\documentclass[pre,twocolumn,english,aps,prb,a4paper,floatfix]{revtex4-1}
\usepackage[T1]{fontenc}
\usepackage{color}
\usepackage[latin1]{inputenc}
\usepackage{graphicx}
\usepackage{bm}
\usepackage{epsfig}

\begin{document}
\title{Confinement of two-dimensional rods in slit pores and square cavities.}

\begin{abstract}
Using Monte Carlo simulation, we analyse the behaviour of two-dimensional hard rods in four different types of geometric confinement: (i) a slit pore where the particles are confined between two parallel walls with homeotropic anchoring; (ii) a hybrid slit pore formed by a planar and a homeotropic wall; square cavities that frustrate the orientational order by imposing either (iii) homeotropic or (iv) planar wall anchoring. We present results for the state diagram as a function of the packing fraction and the degree of confinement. Under extreme confinement, unexpected states appear with lower symmetries than those of the corresponding stable states in bulk, such as the formation of states that break the anchoring constraints or the symmetry imposed by the surfaces. In both types of square cavities the particles form disclinations at intermediate densities. At high densities, however, the elastic stress is relaxed via the formation of domain walls where the director rotates abruptly by ninety degrees. 
\end{abstract}
\date{April 20, 2015}

\author{Thomas Geigenfeind}
\affiliation{Theoretische Physik II, Physikalisches Institut,
  Universit{\"a}t Bayreuth, D-95440 Bayreuth, Germany}
\author{Sebastian Rosenzweig}
\author{Matthias Schmidt}
\affiliation{Theoretische Physik II, Physikalisches Institut,
  Universit{\"a}t Bayreuth, D-95440 Bayreuth, Germany}
\author{Daniel de las Heras}
\email{delasheras.daniel@gmail.com}
\affiliation{Theoretische Physik II, Physikalisches Institut,
  Universit{\"a}t Bayreuth, D-95440 Bayreuth, Germany}

\maketitle

\section{Introduction}

A wealth of new phenomena arises when a liquid crystal is confined in a pore, even in the simplest of geometries where a slit pore is formed by two identical walls that are parallel to each other. In that case the isotropic-nematic transition inside the pore is shifted with respect to the bulk phase transition and terminates in a capillary critical point at a specific value of the wall separation distance. The capillary binodal or capillary nematization line \cite{doi:10.1080/002689797170716,PhysRevE.63.051703} forms the analogue of the capillary condensation line in simple fluids. When confining a smectic phase, layering transitions \cite{PhysRevLett.94.017801} and the suppression of the nematic-smectic transition for specific pore widths \cite{PhysRevE.74.011709} occur due to a commensuration effect between the size of the pore and the smectic layer spacing. In a hybrid pore, formed by two parallel walls with two antagonistic anchoring conditions \cite{barbero1983critical,PM,Galabova,PhysRevE.79.011712}, a balance between the anchoring strengths of the two walls and the elastic energy of the liquid crystal determines the director configuration. The director field can either gradually rotate or generate a step-like defect. In the latter, the director rotates abruptly by $90^{\text{o}}$. If one surface imposes much stronger anchoring than the other, then the director can be approximately constant in the entire capillary. Disclinations are another genuine effect of confinement of liquid crystals. Disclinations can appear by curvature of the underlying space \cite{Lubensky}, e.g. by confining a liquid crystal on the surface of a sphere \cite{Lowen,Bowick,PhysRevLett.108.057801,C3SM52394B}.  The geometry and the dimensionality of the system restrict the topology of the defects that can arise in a given system. The equilibrium director configuration can be highly non-trivial as a result of the balance between the inner forces of the system and the external interaction with the surfaces.

In this paper we study the corresponding two-dimensional system of confined liquid crystals. We investigate whether the above phenomenology of confinement effects persists in two dimensions. We model the particles as hard rods with rectangular shape. This model has been previously investigated. The bulk phase behaviour was analyzed with Monte Carlo (MC) simulation and density functional theory (DFT) ~\cite{SCHLACKEN,raton064903,PhysRevE.80.011707,PhysRevE.77.011707}. The confinement in a planar slit pore, i.e. a slit pore in which both walls promote planar anchoring, has been studied with DFT~\cite{Yuri} using the restricted-orientation approximation (the orientation of the particles is restricted to two perpendicular axes). Triplett and Fichthorn used orientational-bias MC simulation \cite{PhysRevE.77.011707} to study a planar slit pore with selected system sizes. In ~\cite{Miguel} Gonz\'alez pinto et. al. analysed the confinement in square cavities with planar anchoring using DFT in the restricted-orientation approximation. Finally, very recently, the orientational ordering in circular cavities of selected radii has been analyze with MC simulation~\cite{domainwalls}.

Here, we confine the particles in homeotropic and hybrid slit pores and square cavities. We study the phase behaviour by means of Monte Carlo simulation. For all the geometries investigated, we present the state diagram in the plane of system size and density. We find the expected phenomena, such as capillary nematization and smectization and the formation of disclinations in closed cavities. Moreover, we find unexpectedly domain walls and states that break the anchoring or the symmetry imposed by the surfaces. We explain qualitatively the stability of these new states as a balance between different contributions to the free-energy. We expect these new states to appear also in three-dimensional systems under extreme confinement, a regime that as not been investigated yet.

Despite the simplicity of the model, the results can be relevant to understand a variety of systems where packing plays a crucial role such as e.g. experiments on vibrated granular rods \cite{Galanis2010a,indios,indios1}, the adsorption of colloids on substrates \cite{PhysRevLett.109.108303}, the confinement of actin filaments \cite{Mulder} and colloids \cite{C4SM01123F} in quasi-two-dimensional geometries, the assembly of anisotropic nanoparticles at liquid-liquid \cite{Lucio} or liquid-air \cite{kim2001langmuir} interfaces, and the confinement of magnetic nanorods~\cite{magneticrods}.

\section{Model and simulation method}
\label{model}
We consider a system hard rectangular particles of length-to-width ratio $L/D=20$ that interact through excluded volume interactions. The position vector $\vec r_i$ of the center of mass of the $i$-th particle and the unit vector along the long particle axis $\hat u_i$ determine the configuration of rod $i$. We confine $N$ of such particles between two parallel walls or in a square cavity. The interaction between the rods and the surfaces is modelled via an external hard potential $v_{\hbox{\tiny ext}}(\vec r,\hat u)$. We use hard walls (all corners of a particle cannot penetrate the wall) to promote planar anchoring such that the long particle axis aligns preferentially parallel to the walls. We use hard center-walls (the center of mass of a particle cannot penetrate the wall) to promote homeotropic anchoring, that is, the long particle axis aligns preferentially perpendicular to the wall:
\begin{eqnarray}
\beta v_{\hbox{\tiny ext}}(\vec r,\hat u)=\left\{\begin{array}{cl}\infty,&
\hbox{\small at least one corner outside} \\ &\hbox{\small wall (planar)}\\
0,&\hbox{all corners inside the system,}\end{array}\right.
\end{eqnarray}
\begin{eqnarray}
\beta v_{\hbox{\tiny ext}}(\vec r,\hat u)=\left\{\begin{array}{cl}\infty,&
\hbox{\small center of particle outside wall} \\ &\hbox{\small (homeotropic)}\\
0,&\hbox{center of particle inside the system,}\end{array}\right.
\end{eqnarray}
where $\beta=1/k_BT$, with $k_B$ the Boltzmann constant and $T$ the absolute temperature. A hard wall induces planar anchoring even for a single particle because a rod sufficiently close to the wall must adopt a planar configuration. In contrast, homeotropic alignment at a hard center wall emerges from collective behaviour (a single particle sufficiently close to the wall can adopt any orientation). The homeotropic alignment promoted by hard center walls have been previously shown in MC simulation \cite{doi:10.1080/00268979909483083,Lowen} and DFT studies in two \cite{PhysRevE.79.061703} and three \cite{heras:4949,0953-8984-19-32-326103} dimensions using different particle shapes. Given the different origin of both types of anchoring, we expect the planar wall to promote stronger alignment of the particles than the homeotropic wall.

\begin{figure}
\includegraphics[width=3.2in,angle=0]{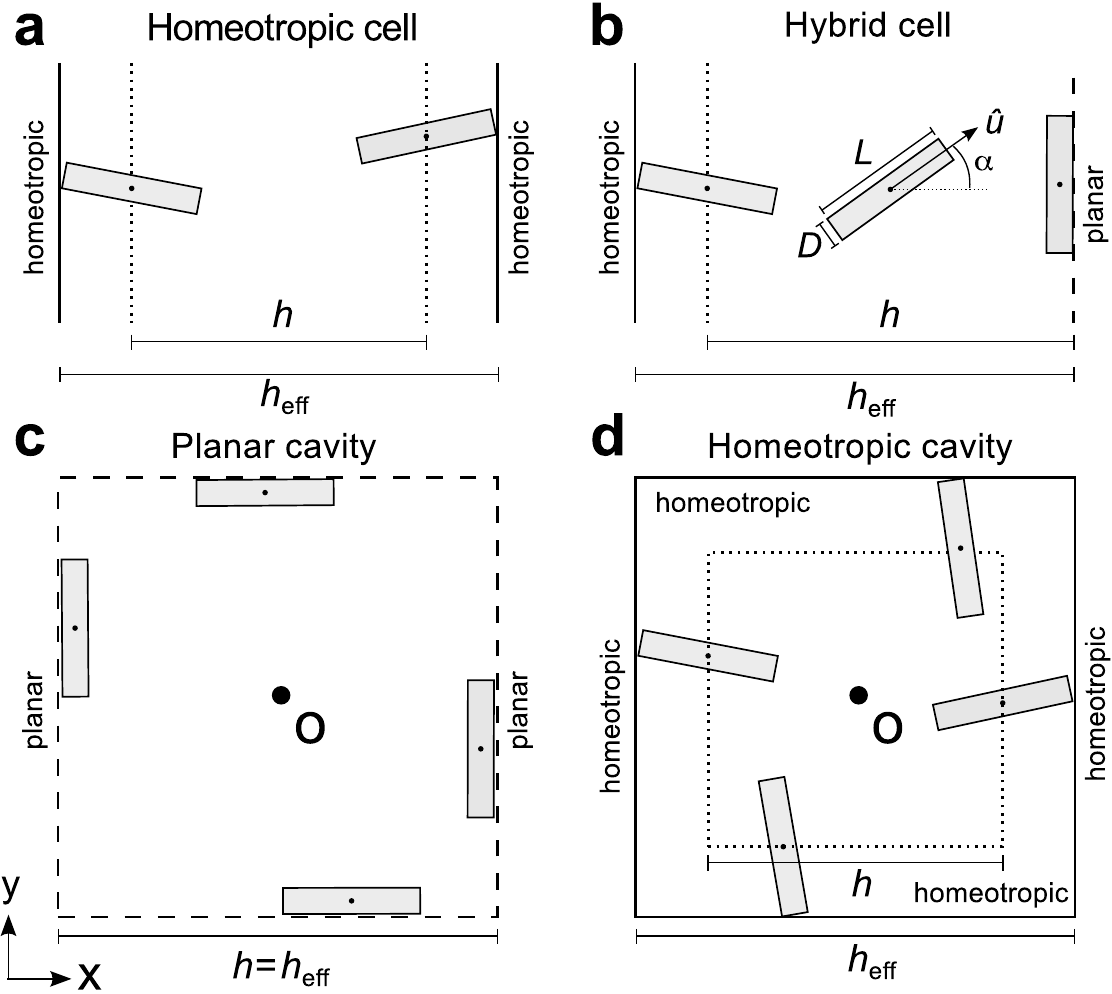}
\caption{Schematic of the different geometries analysed. (a) Homeotropic cell. (b) Hybrid cell. (c) Planar      square cavity. (d) Homeotropic square cavity. Periodic boundary conditions in the y-axis are used in both (a) and (b). In all cases the origin of coordinates is located in the middle of the simulation box. Dotted lines represent hard center-walls, i.e., the center of masses of the particles cannot penetrate the wall (homeotropic anchoring). Dashed lines indicate hard walls, i.e., the corners of the particles cannot penetrate the wall (planar anchoring). Solid lines represent the effective walls in the case of homeotropic surfaces. $h$ is the distance between two parallel walls and $h_{\text{eff}}$ is the effective distance that accounts for the extra space that can be occupied by the corners of the particles if homeotropic walls are present. In the middle of the hybrid cell (b) we show the particle geometry: a hard rectangle of length $L$ and width $D$. The unit vector along the main axis is $\hat u$ and $\alpha$ is the angle between $\hat u$ and the horizontal axis.}
\label{fig1}
\end{figure}

The different geometries that we have analysed are schematically represented in Fig. \ref{fig1}. In the homeotropic cell (Fig. \ref{fig1}a) the rods are confined between two parallel hard center-walls on the $x$ axis. We set periodic boundary conditions in the $y$-direction. The length of the simulation box in the $y$-direction is $h_y=10L$. In the hybrid cell (Fig. \ref{fig1}b) one of the walls promotes homeotropic anchoring and the other induces planar alignment of the particles. Periodic boundary conditions are again applied along the $y$ axis. We also study confinement in square cavities with planar (Fig. \ref{fig1}c) and with homeotropic (Fig. \ref{fig1}d) walls. In all cases we fix the origin of coordinates in the middle of the simulation box. $h$ designates the side length of the square cavities and the distance between the parallel walls of the homeotropic and hybrid cells. In order to compare between different geometries, we use an effective distance $h_{\text{eff}}$ (see Fig. \ref{fig1}). It takes into account that the center-walls (homeotropic anchoring) can be penetrated by the particles by a distance $\sqrt{L^2+D^2}/2$. Hence, for homeotropic pores and homeotropic square cavities $h_{\text{eff}}=h+\sqrt{L^2+D^2}$, for hybrid pores $h_{\text{eff}}=h+\sqrt{L^2+D^2}/2$, and for planar square cavities $h_{\text{eff}}=h$. We define the packing fraction $\eta$ as the ratio between the area covered by the particles and the total area of the simulation box, i.e. $\eta=NLD/A$ with $A=h_{\text{eff}}^2$ for the square cavities and $A=h_{\text{eff}}h_y$ in the case of slit pores. 

We study the equilibrium configurations of the confined rod systems by means of (standard) MC simulations at fixed number of particles and system area. Temperature is irrelevant for hard core systems. Following the ideas in \cite{domainwalls}, we initialize the system at low densities ($\eta<0.1$) with the particles randomly located and oriented in the simulation box. Then we equilibrate the system and perform $\sim10^6$ Monte Carlo steps (MCS) to obtain equilibrium configurations. Here, a MCS is defined as $N$ single particle trial moves, consisting on translation and rotation. Once the simulation ends we insert a few particles and run a new MC simulation. We repeat the insertion of particles until the desired packing fraction is reached or until no new particles can be added. Finally, in order to rule out metastable states, we repeat the whole process but now starting with the last configuration of the simulation (high packing fraction) and removing a few particles each time. In this way we can compare the configurations obtained by increasing and decreasing the number of particles. They should be the same because gradual transitions between states are expected giving the dimensionality of the system. In order to insert new particles we randomly choose one particle and create a parallel replica displaced by $\sim\pm D$ in the direction of the short particle axis. Then we move and rotate the new particle a few thousands times. When the insertion of a new particle leads to overlap with other particles we choose a new rod to create the replica. When decreasing the number of particles we just select one particle at random and remove it. The number of inserted/removed particles from one simulation to another is chosen such that the change in packing fraction is small, between $\sim2\times10^{-2}$ (at low densities) and $\sim5\times10^{-3}$ (at higher densities). 

The probability $p_{\text{a}}$ of accepting one particle move depends on the maximum displacement $\Delta r_{\text{max}}$ and maximum rotation angle $\Delta\alpha_{\text{max}}$ each particle is allowed to perform in one MCS. We aim for $p_{\text{a}}\sim0.25$. Both $\Delta r_{\text{max}}$ and $\Delta\alpha_{\text{max}}$ are calculated each time we change $N$. For each $N$ we first find $\Delta\alpha_{\text{max}}$ in order to accept half of the rotations. Next we find $\Delta r_{\text{max}}$ in order to achieve the desired value of $p_{\text{a}}$.

We characterize the structural properties of the confined rods with three local fields: the density $\rho(\vec r)$, the uniaxial order parameter $S(\vec r)$ and the tilt angle $\psi(\vec r)$. $S(\vec r)$ is defined as the largest eigenvalue of the local order tensor $Q_{ij}(\vec r)=\langle 2\hat u_i\hat u_j-\delta_{ij}\rangle$, where $\hat u_i=(\cos\alpha_i,\sin\alpha_i)$ is the unit vector along the main axis of the $i-$particle, $\delta$ denotes the Kronecker delta, and $\langle\dots\rangle$ is a canonical and spatial average. The tilt angle $\psi(\vec r)$ is the angle between the local director (given by the eigenvector of $S(\vec r)$) and the $x$ axis. The local quantities $\rho(\vec r)$, $S(\vec r)$ and $\psi(\vec r)$ are defined, at each $\vec r$, as an average over $\sim10^4$ different configurations at intervals of $10^2$ MCS. Due to the symmetry in the case of slit pores, the dependence of the local fields is only on the $x$ axis (the axis perpendicular to the walls). We divide the $x$ axis in $\sim10^2$ equidistant bins and for each bin we calculate the local fields by including all the particles with its center of mass located inside the bin. For the square cavities (Fig. \ref{fig1}c and \ref{fig1}d) we study the full $x$ and $y$ dependence of the local fields. In order to obtain spatially smooth fields we calculate the local order tensor by including, for each $\vec r=(x,y)$, all the particles whose center of mass is located in a circle of radius $0.5L$ centered at $\vec r$.

\section{Results}
\label{Results}
In bulk a fluid of hard rectangles of aspect ratio $L/D=20$ undergoes a phase transition from an isotropic to a nematic phase upon increasing the density \cite{SCHLACKEN,raton064903,PhysRevE.80.011707,PhysRevE.77.011707}. The bulk transition is continuous, presumably of the Kosterlitz-Thouless type \cite{Bates,Cuesta,lagomarsino}. The aim of the present work is to study confinement properties, and hence we have not analysed in detail the bulk properties. However we have run bulk simulations using a square box of side length $13L$ with periodic boundary conditions and found a continuous isotropic-nematic transition at $\eta\approx0.27$, in agreement with the predictions of the scale particle theory~\cite{raton064903}. 

In what follows we show the states that occur under confinement. For each geometry we group the states in a diagram as a function of the system size and the packing fraction. We use the local fields (density, uniaxial order parameter and tilt angle) to distinguish between distinct states. However, one should bear in mind that given the confined geometries analyzed here and the dimensionality of the system, one expects gradual transitions between the different states. In addition, the distinction between states, such as nematic, smectic, or isotropic, in confined geometry is not clearcut. 

\subsection{Slab geometry: homeotropic cell}
\begin{figure}
\includegraphics[width=0.95\columnwidth,angle=0]{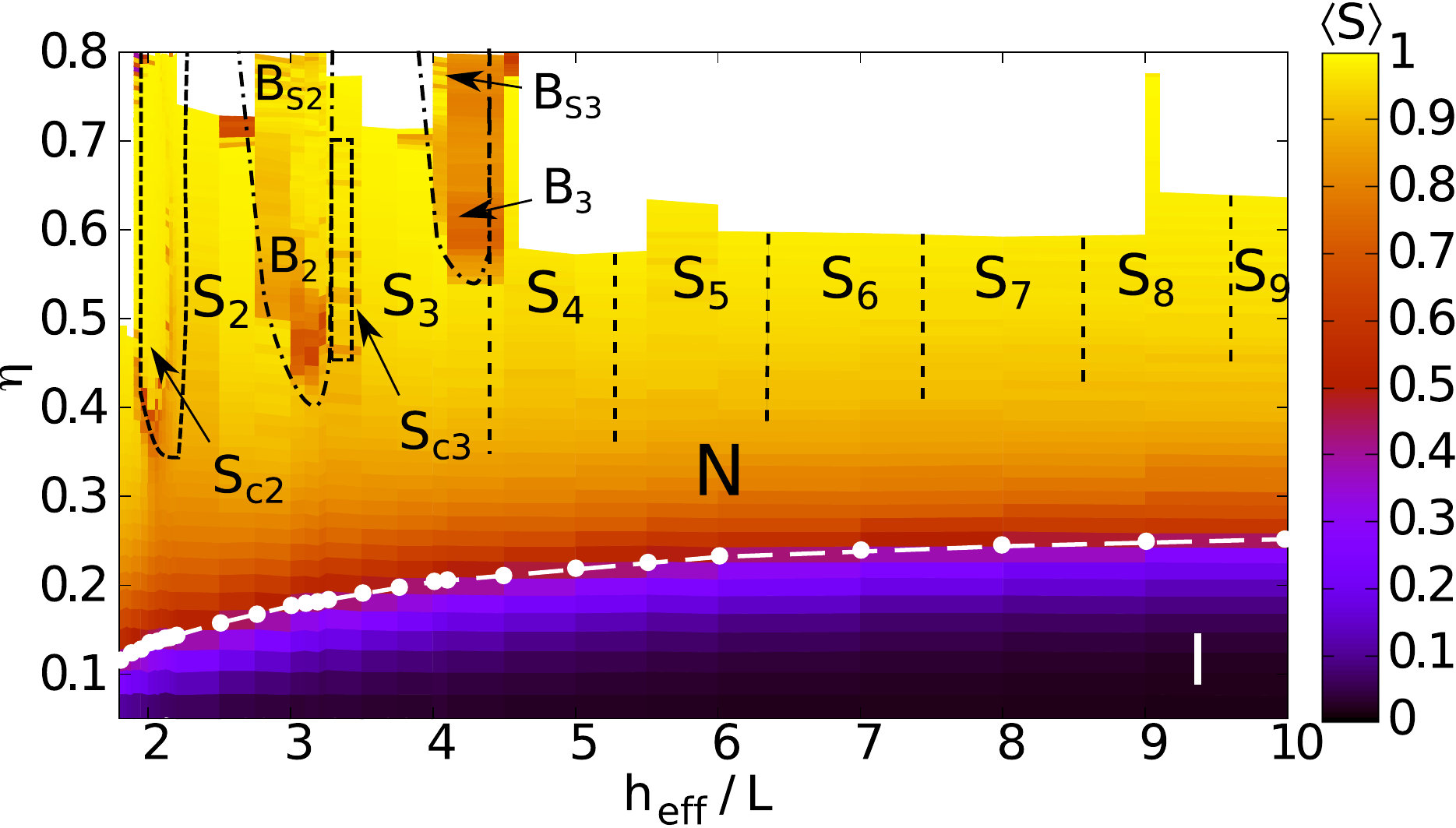}
\caption{State diagram as a function of scaled pore width $h_{\text{eff}}/L$ and packing fraction $\eta$ for hard rectangles ($L/D=20$) confined in a homeotropic cell. The color indicates the average uniaxial order parameter $\langle S \rangle$. White circles show the state points where $\langle S \rangle = 0.5$ (the white-dashed line is a guide to the eye). The black lines delimit approximate boundaries between different states (no simulation data is available for the regions depicted in white at high packing fractions).}
\label{fig2}
\end{figure}
We first consider hard rectangles ($L/D=20$) confined between two parallel planar walls inducing homeotropic anchoring (see Fig. \ref{fig1}a). The state diagram in the plane of packing fraction as a function of pore width is depicted in Fig. \ref{fig2}. We have calculated about $2400$ state points with $h_{\text{eff}}$ varying between $1.8L$ and $10L$. The color map shows the average of the uniaxial order parameter inside the pore
\begin{equation}
\langle S \rangle = \frac1h\int_{\text{pore}}dxS(x).
\end{equation} 

\begin{figure*}
\includegraphics[width=0.85\textwidth,angle=0]{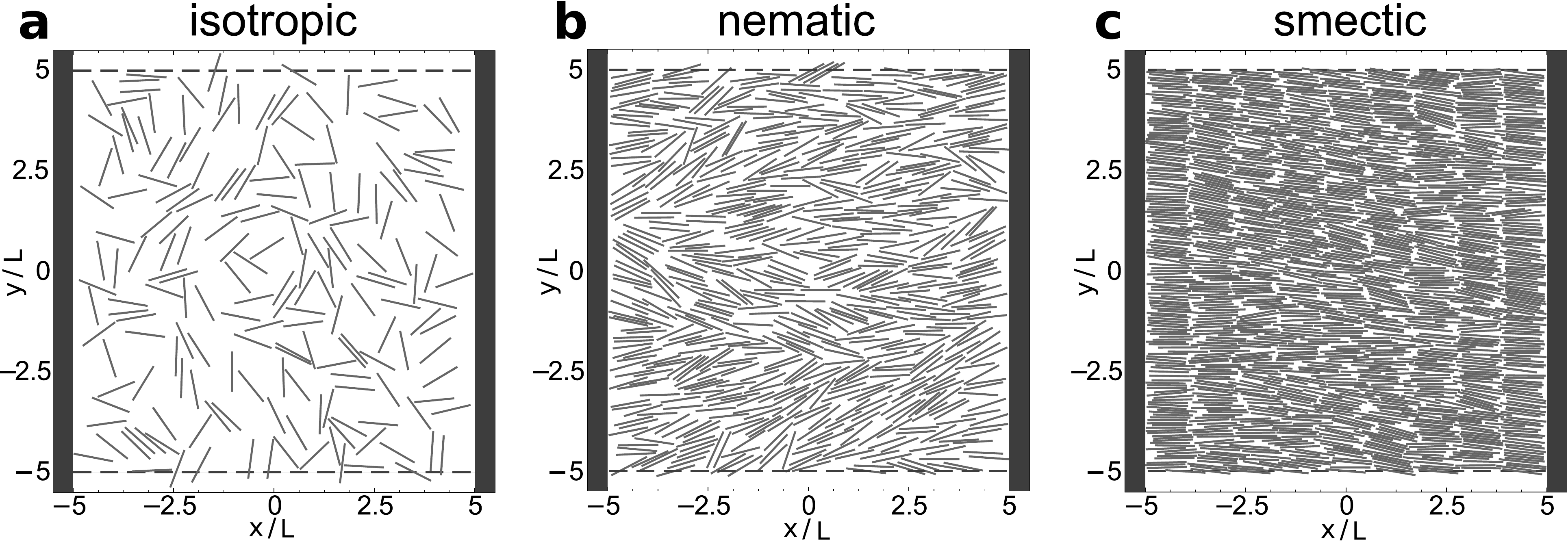}
\caption{Snapshot of characteristic configurations of the particles in a homeotropic slit cell with $h_{\text{eff}}=10L$. The thick vertical lines represent the effective walls. The horizontal dashed lines indicate the location of the periodic boundaries. (a) Isotropic state, $\eta=0.094$. (b) nematic state, $\eta=0.32$. (c) smectic state, $\eta=0.64$. The corresponding order parameter profiles for each state are represented in Fig. \ref{fig4}.}
\label{fig3}
\end{figure*}

\begin{figure}
\includegraphics[width=0.85\columnwidth,angle=0]{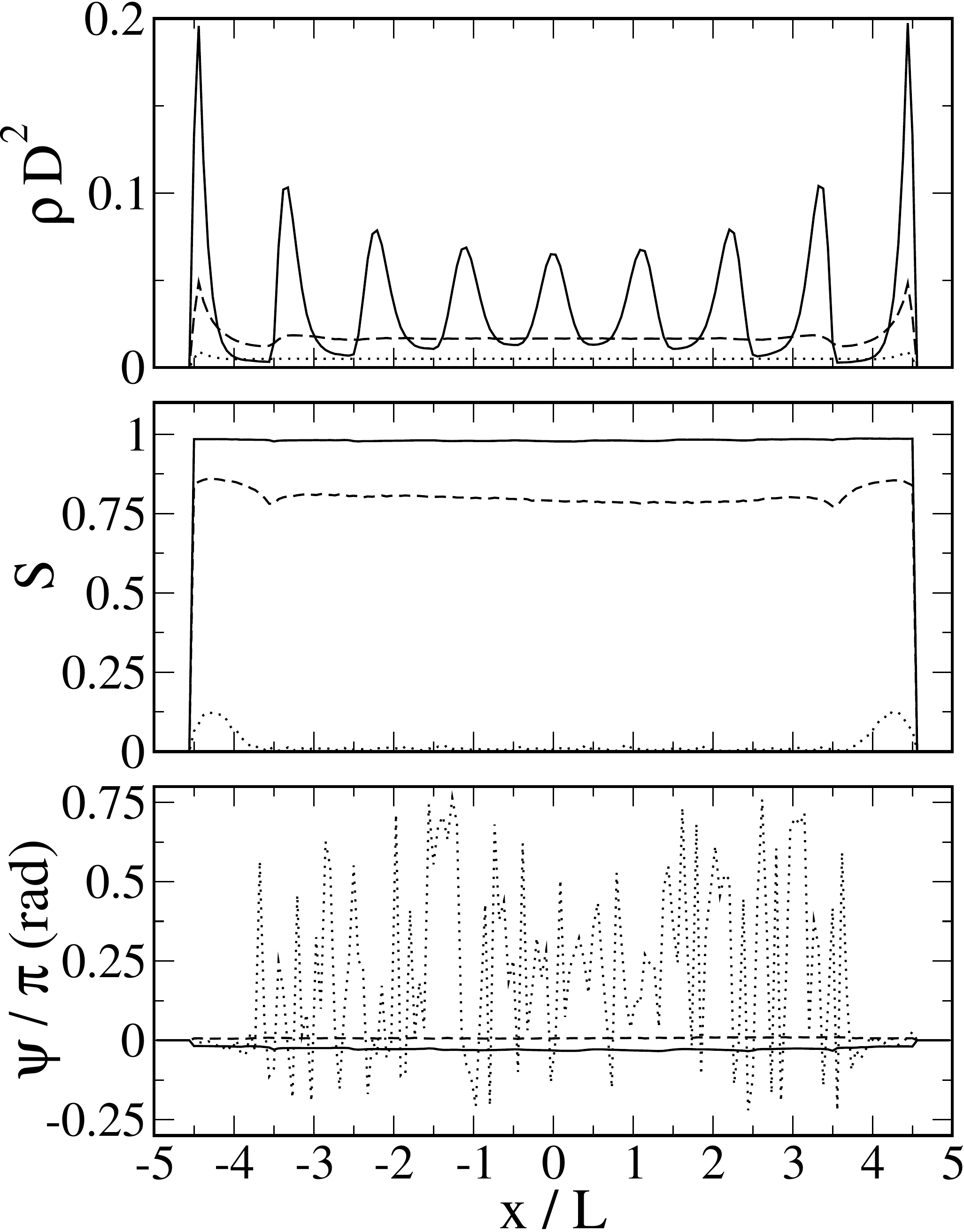}
\caption{Local fields as a function of $x$ of the states in a homeotropic cell with $h_{\text{eff}}=10L$. (top) Scaled density profile $\rho D^2$, (middle) uniaxial order parameter profile $S$ and (bottom) tilt angle profile $\psi$. Dotted line: isotropic state, $\eta=0.094$. Dashed line: nematic state, $\eta=0.32$. Solid line: smectic state (S$_{10}$), $\eta=0.64$. Snapshots of the particle configurations corresponding to these profiles are depicted in Fig. \ref{fig3}.}
\label{fig4}
\end{figure}

\noindent{\bf Isotropic, nematic and smectic states.} First we focus on the larger pores that we have investigated, $h_{\text{eff}}/L\sim10$. We identified three distinct states: isotropic (I), nematic (N), and smectic (S$_i$, with $i$ the number of smectic layers inside the pore). Examples of the particle configurations and the density profiles for each state in a pore with $h_{\text{eff}}=10L$ are shown in Figs. \ref{fig3} and \ref{fig4}, respectively. At low densities the particles form an isotropic state (Fig. \ref{fig3}a). The density profile (Fig. \ref{fig4}, top panel) is rather constant with a small amount of adsorption of particles close to the walls. The uniaxial order parameter (middle panel) is zero except in a small region near the walls where it shows incipient orientational order due to the walls. In this region the particles are (slightly) aligned with their long axes perpendicular to the walls, as the tilt angle (bottom panel) shows. The maximum in density occurs at contact with the surfaces. However, the maximum of the uniaxial order parameter is shifted $0.5-1L$ away from the walls. This is a general feature of a hard center-wall that has been previously reported in three-~\cite{doi:10.1080/00268979909483083,heras:4949,0953-8984-19-32-326103} and two-~\cite{PhysRevE.79.061703} dimensional systems on the basis of MC simulation and DFT. By increasing the density, the capillary nematization (i.e. the formation of a nematic state inside the pore) occurs in a continuous fashion, as expected. In the nematic state all the particles are oriented, on average, perpendicular to the walls (Fig. \ref{fig3}b, and Fig. \ref{fig4} bottom panel). The uniaxial order parameter is positive in the whole capillary (Fig. \ref{fig4} middle)  and there is incipient positional order that propagates into the pore from the walls (Fig. \ref{fig4} top). By further increasing the density the particles form a smectic state (Fig. \ref{fig3}c) with well-defined layers (Fig. \ref{fig4}a), the number of which is the result of commensuration between the size of the pore and the smectic period. In the range of pore sizes investigated here, we have found smectic states with $2$ to $9$ layers. The smectic layers are slightly tilted, especially those in the center of the pore (see the tilt angle profile in Fig. \ref{fig4}c). The reason is that the commensuration is not perfect, i.e. the ratio between the pore size and the smectic period is not an integer number. 

By reducing the size of the pore, the nematization occurs at lower packing fractions. In order to visualize this effect we have depicted a line of constant average uniaxial order parameter, $\langle S \rangle=0.5$, in the state diagram (Fig. \ref{fig2}). This line monotonically increases with $h_{\text{eff}}$ and asymptotically tends to the bulk packing fraction at which $\langle S \rangle=0.5$. Therefore, confinement promotes nematic order. This is an expected behaviour because even at low densities the walls induce some homeotropic anchoring. Similarly, confinement promotes smectic order as oscillations in the density profile start to appear at lower packing fractions in smaller pores.

The most interesting phenomenology arises when the particles are strongly confined in narrow pores and at high packing fractions. In this regime new states with symmetries different than those of the stable bulk phases appear. 

\begin{figure}
\includegraphics[width=0.85\columnwidth,angle=0]{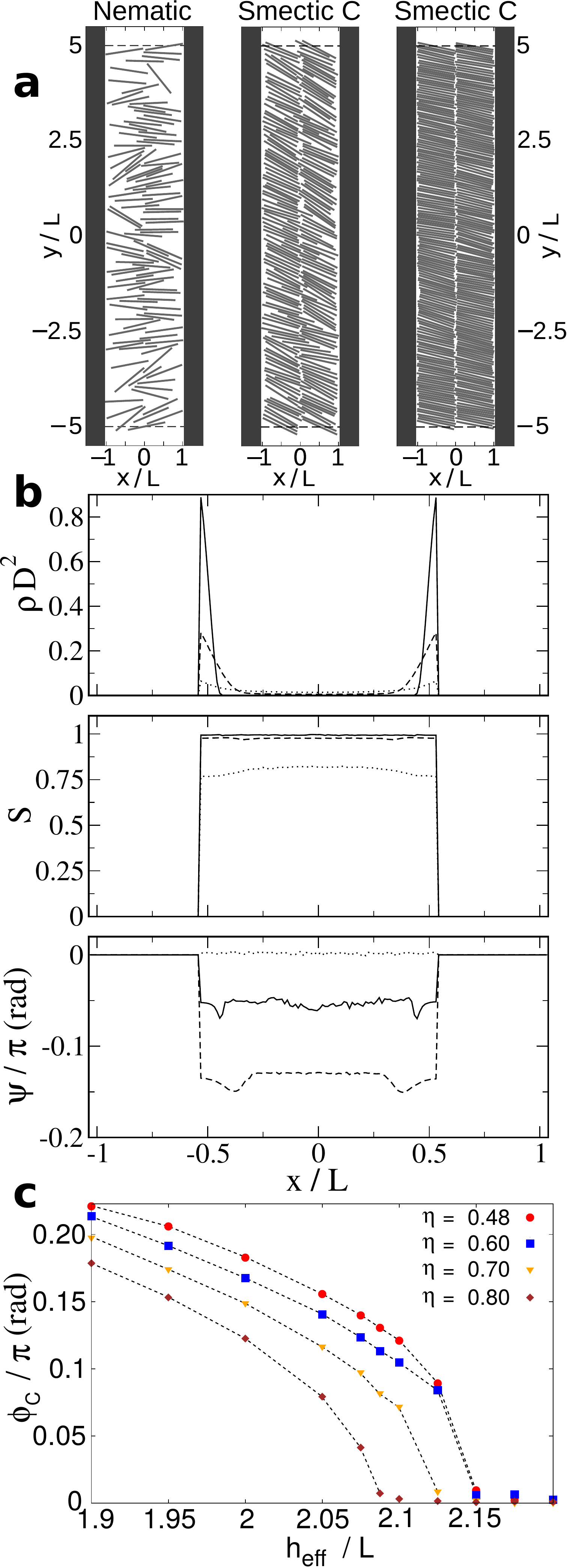}
\caption{(a) Snapshot of characteristic configurations in a homeotropic cell with $h_{\text{eff}}=2.075L$ increasing the number of particles: $\eta=0.25$, nematic state (left); $\eta=0.52$, smectic C state (middle); $\eta=0.79$, smectic C state (right). (b) Local fields as a function of $x$: density profile (top), uniaxial order parameter (middle), and tilt angle (bottom). The different sets correspond to the states showed in panel (a): $\eta=0.25$, nematic state (dotted line); $\eta=0.52$, smectic C state (dashed line); $\eta=0.79$, smectic C state (solid line). (c) Smectic C tilt angle (absolute value) as a function of the pore width for different packing fractions, as indicated.}
\label{fig5}
\end{figure}

\noindent{\bf Smectic C}. For pores with $h_{\text{eff}}\approx2L$ the particles form a nematic state at intermediate densities. By further increasing the density, the rods align into two well-defined layers as already present in the S$_2$ state. Here, however, the particles are strongly tilted with respect to the direction perpendicular to the layers. We call this the smectic C state, S$_{\text{C}2}$, where $2$ indicates the number of layers. The particle configurations and the order parameter profiles of the S$_{\text{C}2}$ state are depicted in Fig. \ref{fig5} panels (a) and (b), respectively. The tilt profile (bottom of panel b) presents two minima shifted from the location of the the maximum density. The smectic C state appears because the size of the pore is not commensurate with the smectic period and the number of layers is reduced. As a result the particles tilt in order to fill efficiently the available space. By increasing the density, the tilt angle decreases. This is consistent with the fact that the smectic period monotonically decreases with the density. In Fig. \ref{fig5}c) we plot the tilt angle at contact with the wall, $\phi_{\text{C}}$, as a function of $h_{\text{eff}}$. For each packing fraction, above a certain threshold, there is a critical pore size below which the particles start to tilt. The smaller the pore is the more tilted the particles are. 

We have also found a smectic C state with three layers, S$_{\text{C3}}$, in a small region of the state diagram around $h_{\text{eff}}=3.25L$ (see Fig. \ref{fig2}). This region is significantly smaller than the stability region of the S$_{\text{C2}}$ state, and it appears at higher packing fractions than $S_{\text{C2}}$ does. At very high packing fractions the smectic period is sufficiently small such that three non-tilted layers fit inside the capillary and the S$_{\text{C3}}$ state is replaced by a S$_{\text{3}}$ state. 

We have not found smectic C states with more than three layers although we cannot rule out their existence in regions of the state diagram at very high packing fractions. Nevertheless, we are confident that those regions shrink rapidly by increasing the pore size and eventually might cease to exist. This can be understood as follows. There is a minimum $d_{\text{min}}$ and a maximum $d_{\text{max}}$ layer spacing between which the formation of non-tilted smectic layers is stable. Consider a capillary with $n$ smectic layers inside. As a rough estimate the layers will tilt if there is no sufficient space to accommodate $n$ non-tilted layers, i.e. if the condition $h_{\text{eff}}/n\le d_{\text{min}}$ is satisfied. In addition $h_{\text{eff}}/(n-1)\ge d_{\text{max}}$ should hold as well, because otherwise an S$_{n-1}$ state, and not an S$_{\text{C}n}$ state, would be stable. Both conditions together roughly set the limits in pore size for a smectic C phase with $n$ layers as
\begin{equation}
(n-1)d_{\text{max}}\le h_{\text{eff}}\le nd_{\text{min}}.\label{minmax}
\end{equation}
The above equation also shows that the range in pore size in which the smectic C is stable decreases with increasing $n$. Indeed, there is a maximum pore size above which no tilted smectic is expected:
\begin{equation} 
h_{\text{eff}}^{\text{max}}=d_{\text{max}}d_{\text{min}}/(d_{\text{max}}-d_{\text{min}}),\label{pinpanpum}
\end{equation}
which results from taking the equality on both sides of Eq. (\ref{minmax}). For hard rectangles $d_{\text{min}}\approx L$ (because the particles cannot overlap). We can express $d_{\text{max}}=d_{\text{min}}(1+\Delta)$, with $\Delta$ being the maximum expansion of the layer spacing for the smectic state in units of $d_{\text{min}}$. Then, using Eq. (\ref{pinpanpum}), $h_{\text{eff}}^{\text{max}}=L(1+\Delta)/\Delta$. We did not find Sm$_{\text{C}}$ states for $h_{\text{eff}}\gtrsim3.25L$. It implies $\Delta\approx0.4$, which seems to be a reasonable value.

\begin{figure}
\includegraphics[width=0.85\columnwidth,angle=0]{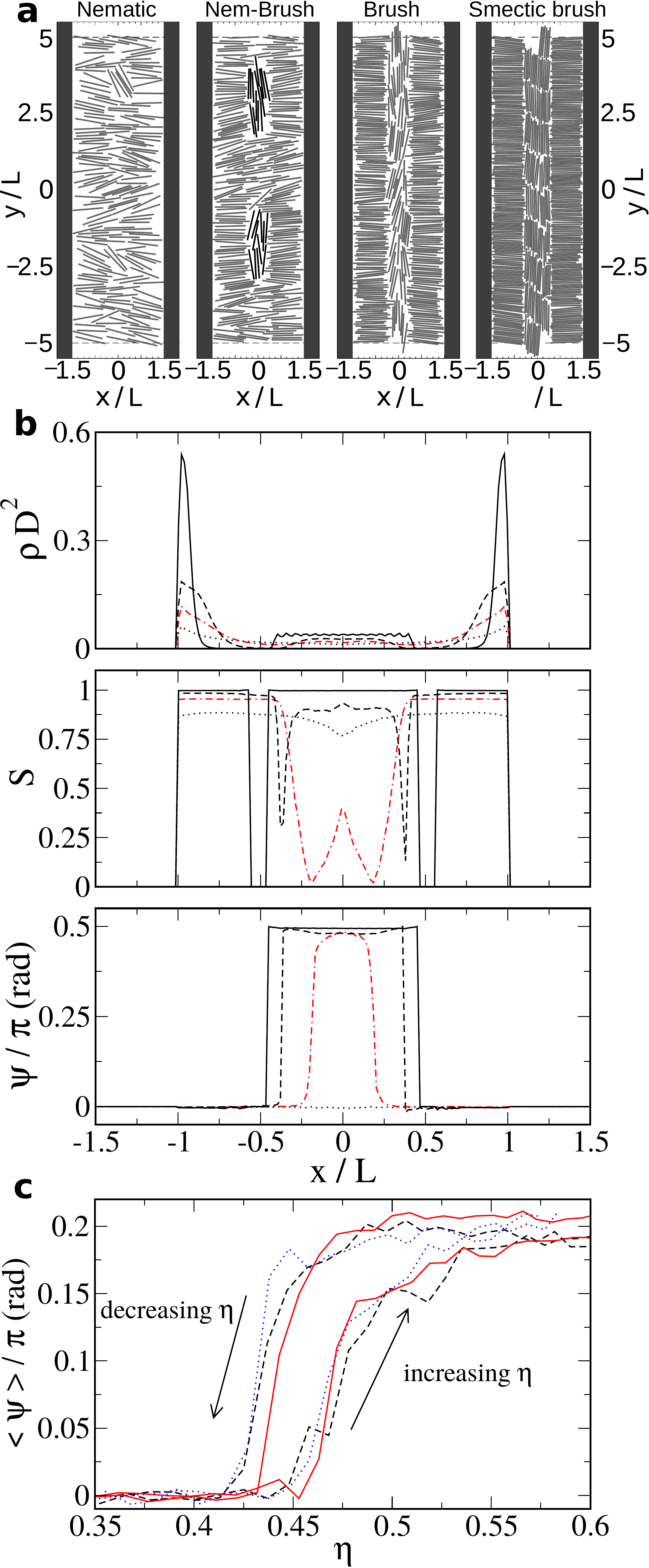}
\caption{Homeotropic cell with $h_{\text{eff}}=3L$. (a) Snapshot of characteristic configurations: nematic state, $\eta=0.31$ (left); nematic-brush, $\eta=0.46$ (second); brush, $\eta=0.59$ (third); smectic-brush, $\eta=0.80$ (right). (b) Local fields as a function of $x/L$ of the states showed in panel (a): density profile (top), uniaxial order parameter (middle), and tilt angle (bottom). The different sets are: nematic (dotted line), nematic-brush (red dotted-dashed line), brush (dashed line), smectic brush (solid line). (c) Average of the tilt angle profile in a pore with $h_{\text{eff}}=3L$ as a function of the packing fraction. Results obtained by increasing and decreasing the number of particles. Simulation parameters: $N_{\text{MCS}}=1.1\times10^6$, $h_y=10L$ (black-dashed line); $N_{\text{MCS}}=1.1\times10^7$, $h_y=10L$ (blue-dotted line); $N_{\text{MCS}}=1.1\times10^6$, $h_y=20L$ (red-solid line). }
\label{fig6}
\end{figure}

\noindent{\bf Brush states.} The remaining states in the diagram depicted in Fig. \ref{fig2} are the brush nematic B$_i$ and the brush smectic B$_{\text{S}i}$, both with $i$ homeotropic layers.

For pores with size in the vicinity of $h_{\text{eff}}\approx3L$ and high packing fractions there is a region where B$_2$ and B$_{\text{S}2}$ are stable. In Fig. \ref{fig6} we show the corresponding order parameter profiles and characteristic particle configurations. By increasing the density from a stable nematic state, some of the particles located in the middle of the pore rotate by ninety degrees, placing their long axis parallel to the walls (nematic-brush state). A further increase in the number of particles results in the pure brush nematic state B$_2$, with one layer of particles with homeotropic anchoring adjacent to each wall. The particles at the center of the cavity are aligned parallel to the walls. To rule out the possibility that this state is an artefact of our method of increasing the number of particles, we have initialized a system with $\eta=0.5$ and $h_{\text{eff}}\approx3L$ in a nematic state with all the particles perpendicular to the walls. After an equilibration stage of about $10^6$ MCS the particles formed the brush state. Hence we are confident that the brush state is indeed stable. The extent of the central region where the particles are oriented parallel to the walls grows by increasing the density (see e.g. the tilt angle profile in Fig.\ref{fig6}, bottom of panel b). In the B$_2$ state those particles in the central region posses orientational but no positional order. However at sufficiently high packing fractions smectic layers occur also at the center. This additional order constitutes the smectic brush state, B$_{\text{S}2}$. 

The transitions between the different states is gradual and hence no differences should appear if we e.g. fix the size of the pore and track the order parameters by first increasing and then decreasing the density. This is actually what we have found for all the states discussed throughout of the paper except for the nematic-brush transition: We plot in Fig. \ref{fig6}c the average of the tilt angle profile,
\begin{equation}
\langle \psi \rangle=\frac1h\int_{\text{pore}}dx\psi(x),
\end{equation}
as a function of the packing fraction, as obtained by either increasing or decreasing the number of particles. $\langle\psi\rangle$ is approximately zero in the nematic state and is different than zero in the brush-nematic state due to the particles aligning perpendicular to the walls. We show in the figure three sets of data corresponding to simulations with different numbers of MCS and different lateral pore sizes. In the three cases there is strong hysteresis, most likely related to a finite size effect because our system can be effectively considered as a one-dimensional system with short range interactions where no first order transitions are expected according to the so-called van Hove theorem. Nevertheless, the van Hove theorem does not apply in the presence of external fields (see \cite{CuestaVH} for a discussion about the exceptions to the van Hove theorem). 

\begin{figure}
\includegraphics[width=0.85\columnwidth,angle=0]{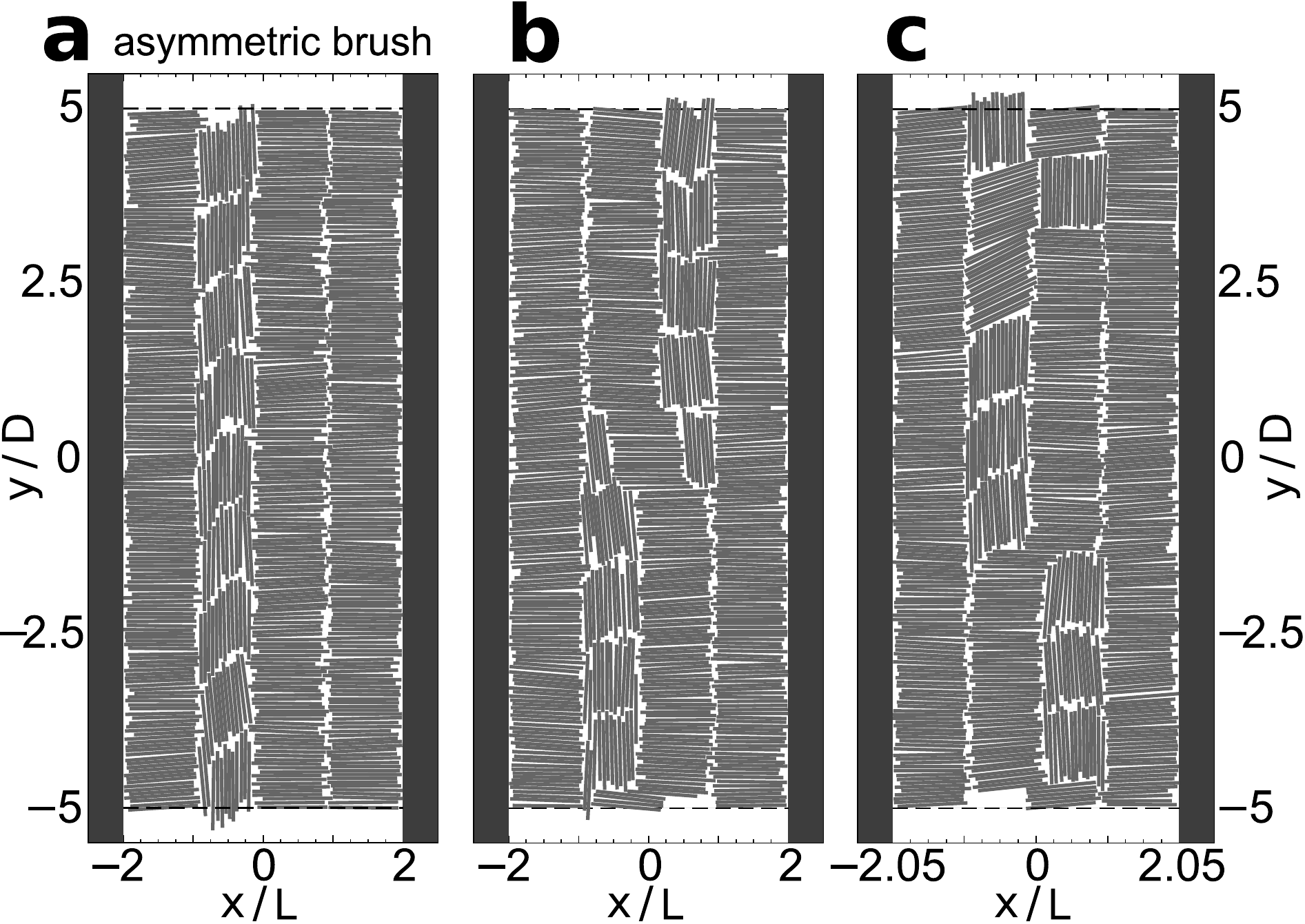}
\caption{Snapshots of the particle configurations in a homeotropic cell with $h_{\text{eff}}=4L$ (a,b) and $h_{\text{eff}}=4.1L$ (c) at an average packing fraction $\eta=0.80$.} 
\label{fig7}
\end{figure}

Brush nematic and smectic states with three layers of particles that are perpendicular to the walls also appear in the region of the state diagram where $h_{\text{eff}}\approx4L$ (see the labels B$_3$ and B$_{\text{S}3}$ in Fig. \ref{fig2}). An example of the B$_{\text{S}3}$ state is shown in Fig. \ref{fig7}a. The $B_3$ and $B_{\text{S}3}$ states are in general not symmetric in $x$. The region of particles aligned parallel to the walls is not located in the middle of the cell. In order to test the stability of the unexpected symmetry breaking of these states, we have initialized the system in a symmetric configuration where two small regions of rods parallel to the walls are placed between layers of particles with homeotropic configuration. The resulting configurations after running more than $10^7$ MCS are shown in panels (b) and (c) of Fig. \ref{fig7}. As we had to initialize the system at very high packing fractions we were not able to recover the asymmetric brush profile shown in (a). However, the states in (b) and (c) are again asymmetric. They resemble coexistence states between the state represented in (a) and its mirror image. Hence, we conclude that the symmetric B$_3$ and B$_{\text{S}3}$ phases are not stable. This could have been anticipated because the symmetric state has four interfaces between parallel and perpendicular rods, whereas the asymmetric one contains only two. Asymmetric profiles in symmetric pores have been previously found in three-dimensional mixtures of hard rods \cite{C003448G} and monocomponent and binary mixtures with soft interactions, see e.g. \cite{PhysRevLett.59.98,1.2715934,1.2895747,Ruckenstein201056}. The states shown in (b) and (c) are probably not stable because they have larger interfaces than the one obtained by gradually increasing $N$ (a). In addition in both cases the central layer is distorted, which increases the elastic energy of the system. Nevertheless, due to the finite-time simulations and the finite lateral pore size, the system may show a bimodal behaviour oscillating between states (b) and (c) as if it were a genuine phase transition.

As for the smectic C states, the regions in the state diagram where the brush states are stable move to high packing fractions and shrink with the size of the pore. We could not find brush states above B$_3$ but their existence in narrow regions and high packing fractions cannot be ruled out. 

Instead of forming a brush state, the particles could tilt and form a smectic C state reaching very high packing fractions, similar to those in the brush state. Hence, an interesting question is: why are there regions of the state diagram where the brush states are more stable than the smectic C? To answer this question we compare the excess in free energy of both states with respect to non-distorted nematic or smectic states. In the brush state the excess in free energy is dominated by the two interfaces of perpendicularly aligned particles. In the smectic C state there are two important contributions: the violation of the anchoring imposed by the surfaces and the formation of tilted layers. Both contributions increase by increasing the tilt angle. The tilt angle in the smectic C state, and hence the excess in free energy, increases by reducing the size of the pore. As a consequence the brush states appear replacing the smectic C when the size of pore is reduced. This simple argument explains not only the appearance of the brush states but also their relative position to the S$_{\text{C}}$ in the state diagram, cf. Fig.~\ref{fig2}. 

The regions in the state diagram (Fig. \ref{fig2}) where smectic C and brush states become stable possess a smaller average uniaxial order parameter than the surrounding regions. When these states start to form there are regions in the pore where the particles align according to the incipient state and other regions where the particles remain in the nematic state. In addition, the order parameter profiles of the brush states depend on both the vertical and the horizontal coordinates. However we calculate them only as a function of the horizontal coordinate. Both effects result in an artificially reduced uniaxial order parameter that, on the other hand, is useful to distinguish the boundaries between states in the state diagram.

\subsection{Slab geometry: hybrid cell}

\begin{figure}
\includegraphics[width=0.95\columnwidth,angle=0]{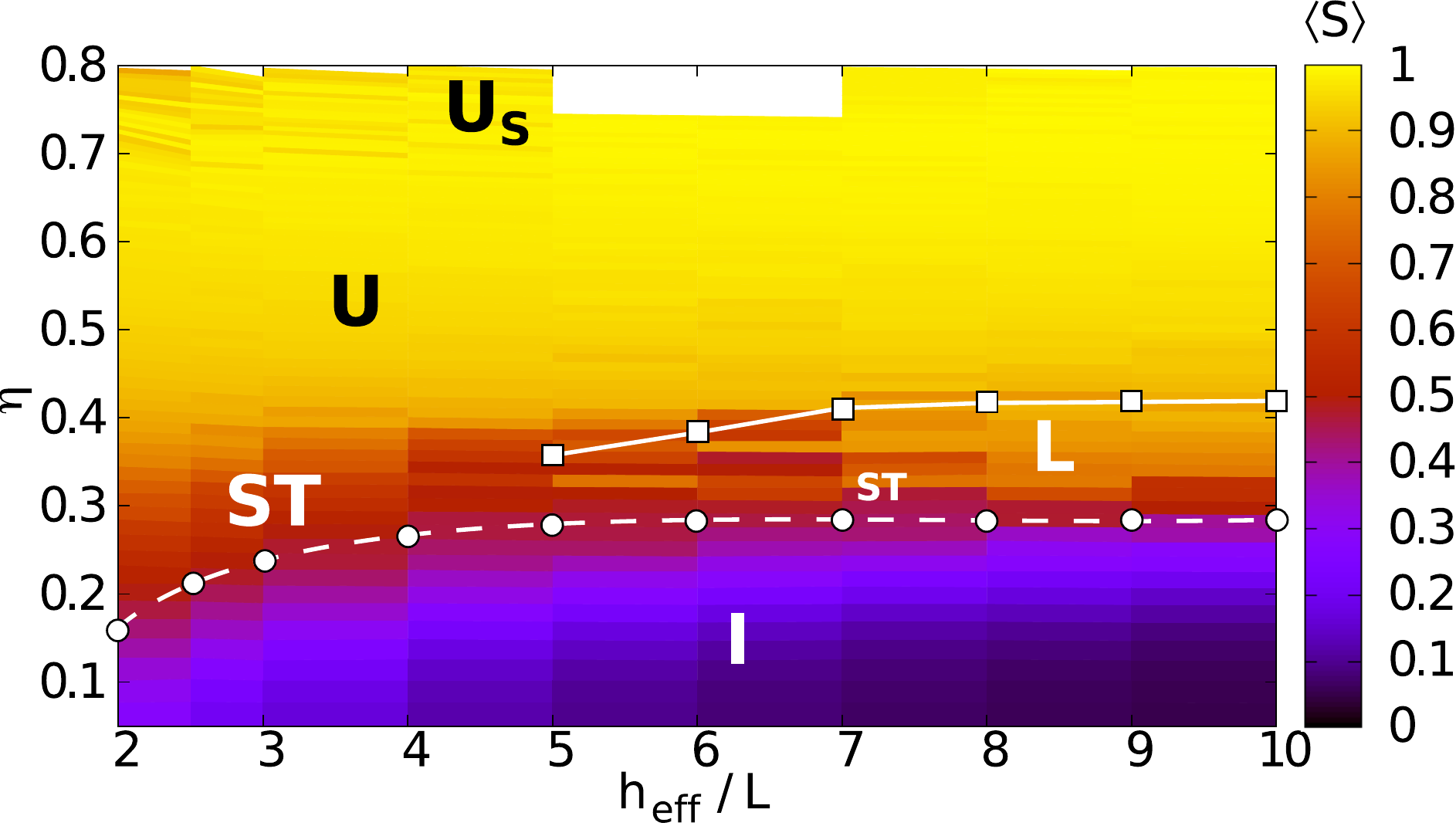}
\caption{State diagram as a function of the scaled pore width $h_{\text{eff}}/L$ and the packing fraction $\eta$ for hard rectangles ($L/D=20$) confined in a hybrid planar cell. The color map represents the value of the averaged uniaxial order parameter $\langle S \rangle$. The empty circles connected with a dashed white line show the state points for which $\langle S \rangle=0.5$. The empty squares connected via a solid white line show the approximate boundary between the linear and the uniform states.}
\label{fig8}
\end{figure}

Next we investigate the behaviour of hard rectangles  with the same aspect ratio $L/D=20$ as before, but confined between two parallel walls that promote antagonistic anchoring, the so-called hybrid cell. The ''left'' wall induces homeotropic anchoring and the ''right'' wall promotes planar alignment of the particles (see a schematic of the geometry in Fig. \ref{fig1}b). The state diagram is depicted in Fig. \ref{fig8} in the plane of packing $\eta$ fraction and scaled-effective pore width $h_{\text{eff}}/L$. As in the previous case the color map indicates the value of the averaged uniaxial order parameter inside the pore. We have run more than $800$ simulations with pore widths $h_{\text{eff}}/L=2-10$ to generate the diagram.   

For any pore width the isotropic state is stable at low densities. In this state there is a small layer of particles oriented perpendicular (parallel) to the left (right) wall. The remaining particles do not show orientational order. As discussed the anchoring imposed by the planar wall is stronger than that of the center-hard wall. A manifestation of this is the the value of uniaxial order parameter in the isotropic state (not shown), which is higher close to the planar wall than close to the homeotropic one. Confinement in a hybrid cell promotes, as in the homeotropic cell,  orientational order of the particles (see e.g. the line of constant uniaxial order parameter in the state diagram).  

\begin{figure*}
\includegraphics[width=1.0\textwidth,angle=0]{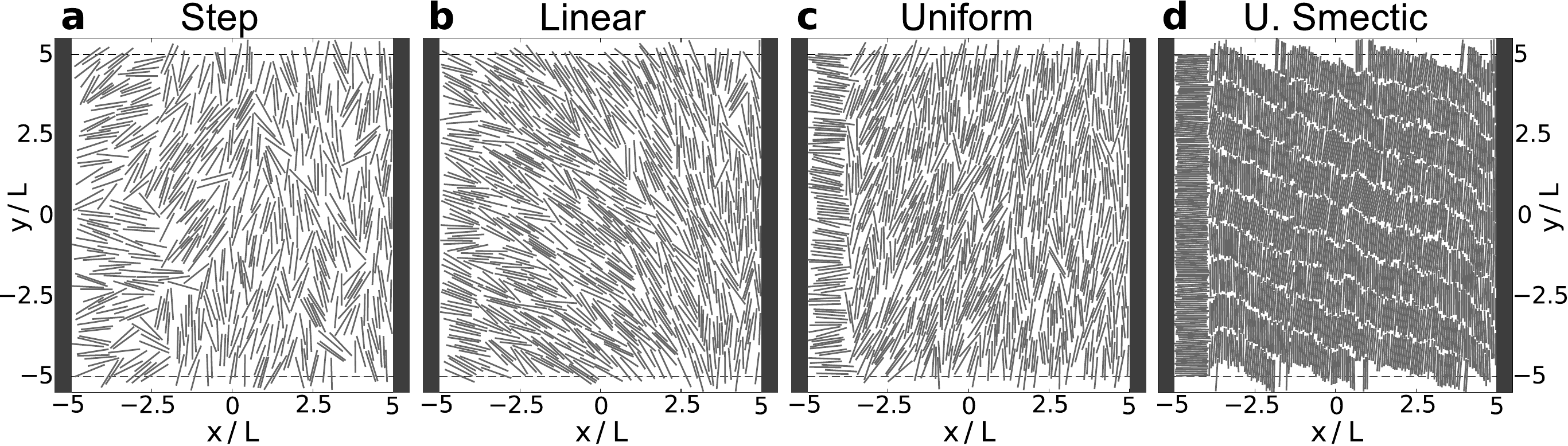}
\caption{Snapshots of a representative configuration of the particles in a hybrid cell with $h_{\text{eff}}=10L$. (a) Step state, $\eta=0.29$. (b) Linear nematic state, $\eta=0.37$. (c) Uniform nematic state, $\eta=0.43$. (d) Uniform smectic state, $\eta=0.78$. The corresponding order parameter profiles of these configurations are shown in Fig. \ref{fig10}. }
\label{fig9}
\end{figure*}

\begin{figure}
\includegraphics[width=0.85\columnwidth,angle=0]{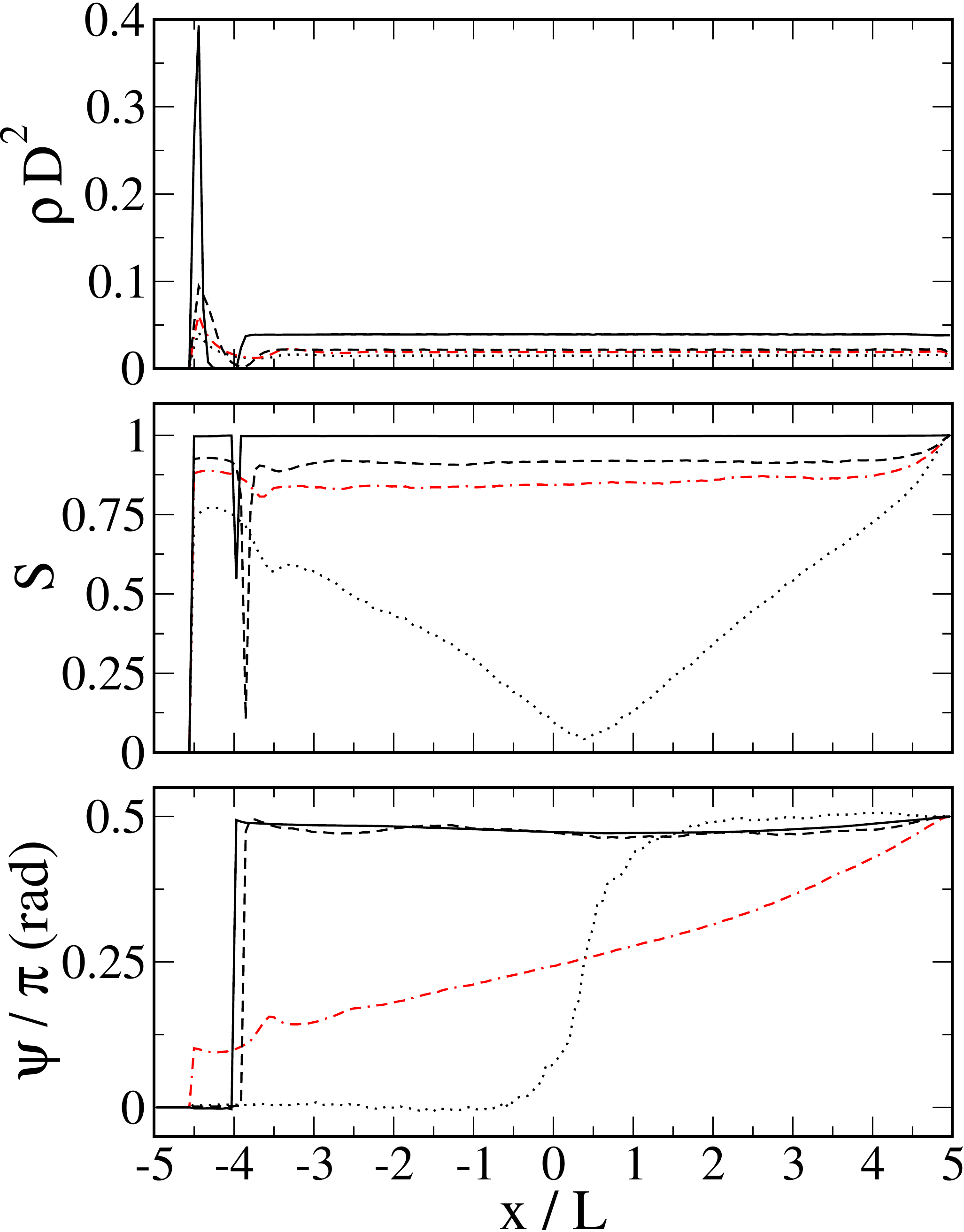}
\caption{Local fields as a function of $x$ of the states in a hybrid cell with $h_{\text{eff}}=10L$. (top) Density profile, (middle) uniaxial order parameter profile, (bottom) tilt angle profile. Dotted line: step state, $\eta=0.29$. Red dotted-dashed line: linear nematic state, $\eta=0.37$. Dashed line: uniform nematic state, $\eta=0.43$. Solid line: uniform smectic state, $\eta=0.78$. Snapshots of the particle configurations corresponding to these profiles are depicted in Fig. \ref{fig9}.}
\label{fig10}
\end{figure}

First we focus on the regime of large pore sizes. By increasing the number of particles the following sequence of states appears: isotropic (I), step (ST), linear (L), uniform nematic (U), and uniform smectic (U$_{\text{S}}$). Examples of the configuration of the particles and the order parameter profiles in the intermediate and high density states for a pore with $h_{\text{eff}}=10L$ are shown in Figs. \ref{fig9} and \ref{fig10}, respectively. 

\noindent{\bf Step state.} Also known as director-exchange phase or biaxial phase, the step phase was proposed by Schopohl and Sluckin \cite{Schopohl} and by Palffy-Muhoray et al. \cite{PM}. It has been study in three-dimensional systems with Landau-de Gennes theory \cite{PM,Sarlah,Galabova}, simulation \cite{paulo,Noe}, and density functional theory \cite{ponce,PhysRevE.79.011712}. In the ST state there are two nematic regions with uniform and opposite directors following the anchoring imposed by the surfaces (see Fig. \ref{fig9}a). The interface between both regions is sharp; the director rotates by ninety degrees in a region of about two molecular lengths (see the tilt profile in Fig. \ref{fig10} dotted line). At the interface the uniaxial order parameter drops to zero. For large pores the ST state is stable in a very narrow region of packing fractions contiguous to the isotropic state. Actually, as suggested in \cite{PhysRevE.79.061703}, the ST state could be a manifestation of the isotropic state at densities close to the capillary nematization in sufficiently narrow pores. 

\noindent{\bf Linear state.} Increasing the packing fraction from the ST state in large pores gives rise to the formation of the linear state (see Fig. \ref{fig9}b). Here the director rotates continuously from homeotropic to planar anchoring (see Fig. \ref{fig10}, red dotted-dashed line). Far enough from both substrates the tilt profile varies linearly with the distance across the pore. In this way the elastic free energy is minimized. The formation of the L state is the analogue to the capillary nematization in a symmetric pore. The L state is compatible with the anchoring imposed by both substrates and at the same time minimizes the elastic free energy. 

\noindent{\bf Uniform nematic state.} By further increasing the packing fraction there is a configurational change from the linear state to the uniform nematic state. The U state is a nematic with uniform director parallel to the wall except for the first layer of particles adsorbed at the homeotropic wall, where the particles are perpendicular to the substrate (see Figs. \ref{fig9}c and \ref{fig10}). This first layer is most likely a consequence of the peculiarities of the hard center-wall, which allows for a high packing fraction of particles only in the case that rods are aligned perpendicular to the wall. The linear-uniform transition is a consequence of the stronger anchoring induced by the planar wall in comparison to the hard-center wall. Although it occurs gradually as we increase the packing fraction, the range in $\eta$ at which the transition occurs is small, enabling us to draw a line in the state diagram that approximately indicates its location (see Fig. \ref{fig2}). 

The density at the L-U configurational change and, therefore, the range in packing fractions at which the L state is stable, increases with the pore width. Actually, the L state may replace the uniform nematic state in the regime of very large pores. To understand this, consider the excess in free energy of the L and U states over a bulk undistorted nematic, $\Delta F_{\text{ex}}$. In the L state $\Delta F_{\text{ex}}^{\text{L}}=F_{\text{A}}^R+F_{\text{A}}^L+F_{\text{el}}$, where $F_A^R$ and $F_A^L$ are due to the anchoring imposed by the right and the left walls, respectively, and $F_{\text{el}}$ is the elastic energy due to the deformations of the director field. For very large pores the director varies linearly and rotates by ninety degrees in the pore. Hence, the divergence of the director is $\nabla\cdot\hat n\approx\pi/(2h)$ and the elastic energy $F_{\text{el}}\approx k_1(\pi/2)^2/h$, with $k_1(\eta)$ the splay elastic constant. In the uniform state both anchoring constraints are satisfied and contribute to the excess in free energy as in the linear state. The director is not distorted ($F_{\text{el}}=0$) but there is an interface generated by the first layer of particles with homeotropic alignment. Hence, $\Delta F_{\text{ex}}^{\text{U}}=F_{\text{A}}^R+F_{\text{A}}^L+F_{\text{I}}$, where $F_{\text{I}}$ is the free energy of the nematic-nematic interface. The elastic contribution in the L state decreases with $h$ but $F_{\text{I}}$ does not. Therefore we expect the L state to replace the U state for sufficiently wide pores. Note that the same argument is valid if instead of an interface between two nematics with opposite directors in the U state there is a violation of the anchoring imposed by one of the substrates. In that case the anchoring energy in the U state would be higher than that in the L state and would not decrease with the size of the pore. We have performed simulations in a pore with $h_{\text{eff}}=20L$, and the L-U transition occurs at $\eta\approx0.6$, considerably higher than e.g. the case $h_{\text{eff}}=10L$ ($\eta\approx0.41$). This scenario, in which the L state replaces the U state in very wide pores, is therefore plausible. Nevertheless, we cannot rule out another scenario in which the U state is stable for any pore width, as it has been found in \cite{PhysRevE.79.061703}. In \cite{PhysRevE.79.061703} a system of spherocylinders confined in a hybrid cell is analyzed with DFT, and the LU transition persists  at any pore length due to an anchoring transition at one of the substrates, i.e. the type of anchoring induced by the wall changes by varying the density. In our case, however, such an anchoring transition is not expected as we deal with hard core potentials. Note, nevertheless, that the first layer of particles adsorbed on the hard-center wall could effectively act as a hard wall for the second layer if the density is sufficiently high, which in practice could be viewed as an anchoring transition. In this second scenario the U state would be stable even for very wide pores. Simulations for pores wider than those considered here could help to elucidate this point.

\noindent{\bf Uniform smectic state.} Finally at very high packing fractions the particles parallel to the walls form smectic layers. The resulting state is similar to the uniform nematic but with positional ordering. An example of the particle configurations is presented in Fig. \ref{fig9}d. The corresponding order parameter profiles are shown in Fig. \ref{fig10} (solid lines). The formation of layers by increasing the density from the U state takes place very gradually and we could not identify the packing fraction of the U-U$_{\text{S}}$ transition in the state diagram. We find that the direction of the layers is not perpendicular to the walls. The director is tilted with respect to the direction perpendicular to the layers, like it is in a smectic C. We did not find any relation between the tilt angle and the size of the pore. The fact that the layers are tilted could be a finite size effect related to the vertical size of the pore, or it could also be related to high fluctuations in the tilt angle.   

In contrast to the homeotropic case, the state diagram of the hybrid cell does not show additional states in the regime of small pores. The only significant difference in the region of small pores with respect to the region of large pores is that the linear state disappears. We could not find the linear state in pores with $h_{\text{eff}\lesssim}5L$.

\subsection{Square cavity with planar walls}\label{sp}

\begin{figure}
\includegraphics[width=0.95\columnwidth,angle=0]{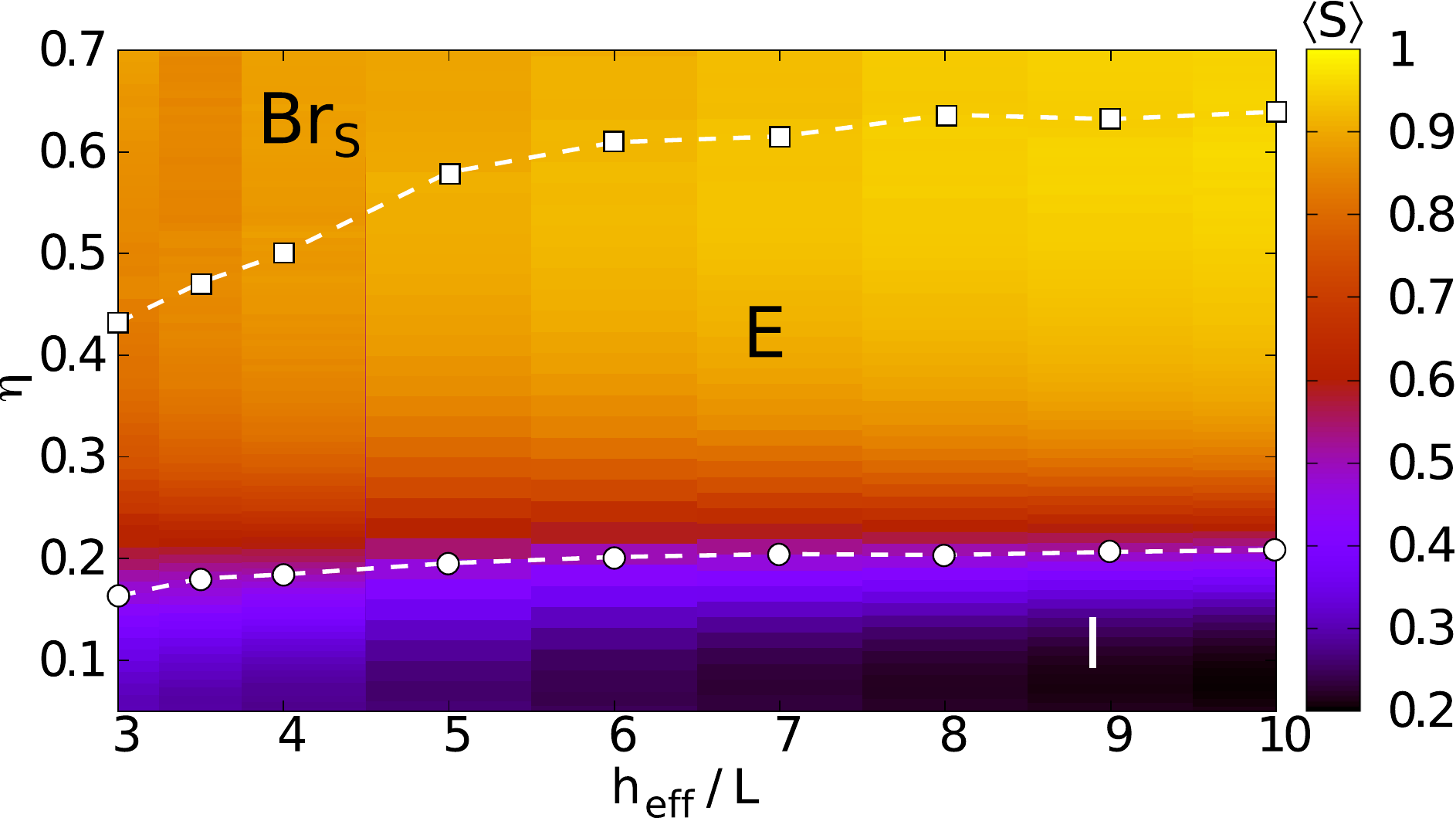}
\caption{State diagram of a fluid of hard rectangles ($L/D=20$) confined in a square cavity with planar anchoring in the packing fraction-side length plane. The color map represents the average of the uniaxial order parameter inside the cavity, $\langle S \rangle$. Empty circles indicate the position where $\langle S \rangle=0.5$. Empty squares roughly show the boundary between the elastic and the bridge smectic states. Lines are guides to the eye.}
\label{fig11}
\end{figure}

\begin{figure*}
\includegraphics[width=0.8\textwidth,angle=0]{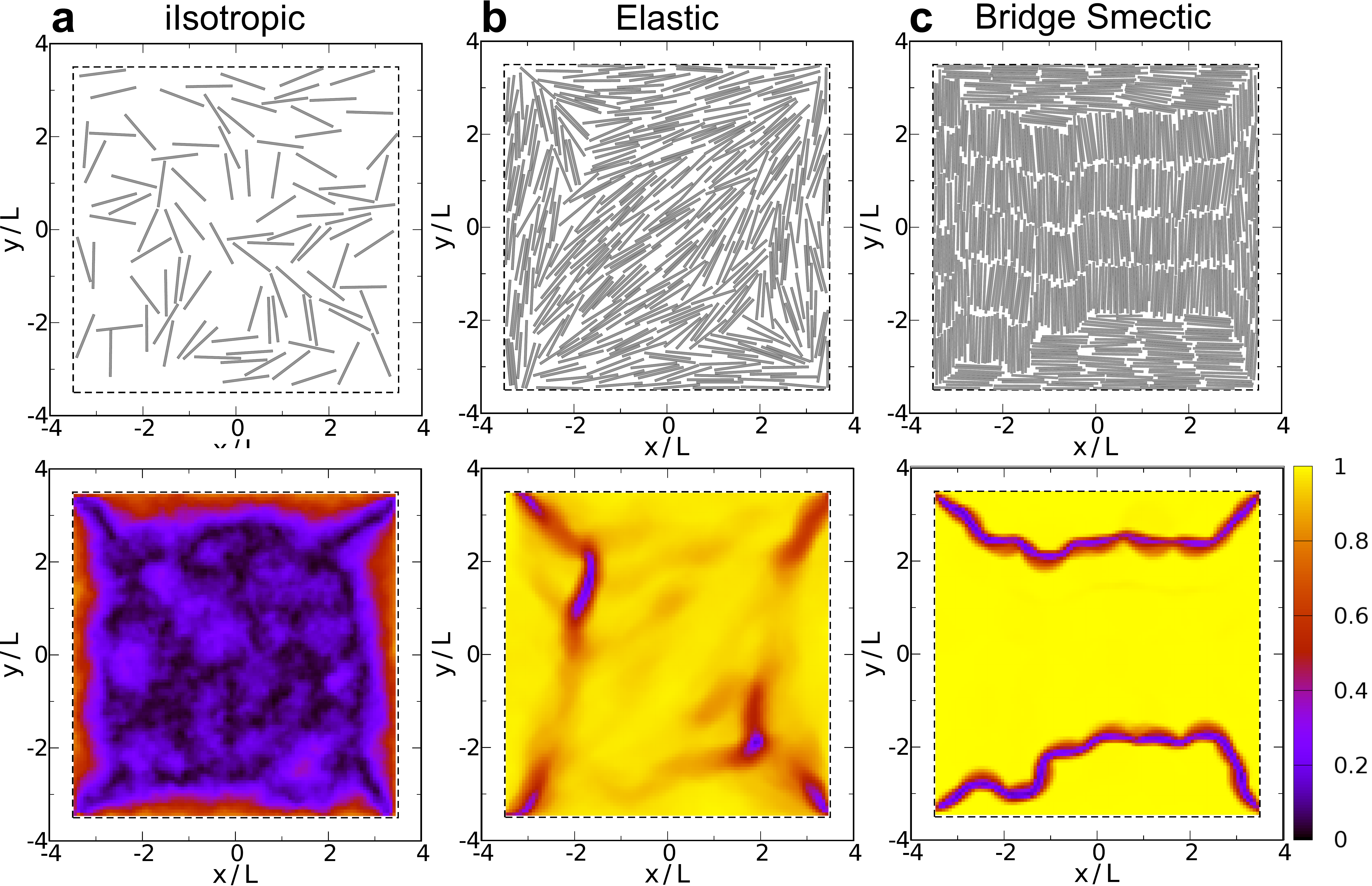}
\caption{Square cavity with planar anchoring and side length $h_{\text{eff}}=7L$. The dashed square indicates the location of the walls. Upper row: snapshots of the particle configurations. Bottom row: local uniaxial order parameter. (a) I state with $\eta\approx0.097$. (b) E state with $\eta\approx0.40$. (c) Br$_{\text{S}}$ state with $\eta\approx0.71$.}
\label{fig12}
\end{figure*}

We next consider confinement of the rods in a square cavity favouring planar alignment of the particles (see a sketch of the geometry in Fig.\ref{fig1}c). Confinement in all spatial directions adds additional constraints on the orientational ordering of the particles that might result in e.g. the formation of topological defects. The state diagram in the plane of packing fraction and side length is depicted in Fig. \ref{fig11}. We found three distinct states: isotropic (I), elastic (E), and bridge smectic (Br$_{\text{S}}$). Representative results of these states are shown in Fig. \ref{fig12} for a cavity with side length $h_{\text{eff}}=7L$.

At low densities the isotropic state is stable, here the fluid is disordered. Only a thin layer of particles close to the walls shows some degree of orientational order (see the uniaxial order parameter in Fig. \ref{fig12}a). The density (not shown) is rather uniform in the whole cavity, showing only a small desorption of particles close to walls, especially near the corners of the cavity. The uniaxial order parameter is also smaller in the vicinity of the corners. The nematization occurs by increasing the number of particles. The result is a gradual transition from the isotropic to the elastic state (see Fig. \ref{fig12}b). In the E state the nematic cannot adopt a uniform configuration due to the surfaces and six disclinations arise in the cavity. Four disclinations are located in the corners of the cavity. In the middle of the cavity the rods align along one of the diagonals. This leads to the formation of two further disclinations with topological charge $-1/2$ located along the other diagonal, at a distance of about $2.5L$ from the corners. The disclinations are clearly visible as a drop of the uniaxial order parameter (see Fig. \ref{fig12}b). The density profile (not shown) also reveals a depletion of particles close the defect cores. The position of the cores of the $-1/2$ defects fluctuates during the simulation but they always stay away from each other. The inner $-1/2$ defects are connected with the adjacent corner defects, see the uniaxial order parameter in Fig.~\ref{fig12}b). The smaller the cavity becomes the stronger this effect is. 

The packing fraction at which the capillary nematization occurs increases monotonically with the size of the cavity and tends asymptotically to the bulk value (see for example the line of constant average uniaxial order parameter depicted in the state diagram, Fig. \ref{fig11}). The average order parameter $\langle S \rangle$ depends nontrivially on $h_{\text{eff}}$ and $\eta$. In the I region, e.g., $\eta=0.1$, $\langle S \rangle$ decreases by increasing $h_{\text{eff}}$ because the walls induce order in a small region close to them and the ratio between this region and the whole cavity decreases as $h_{\text{eff}}$ is increased. Once the nematic is formed the trend is reversed. For instance, at $\eta=0.3$ the smaller the cavity becomes the lower $\langle S \rangle$ is. Here the whole cavity is in a nematic state, except in those regions where the disclinations appear, and the ratio between the surface occupied by the disclinations and the whole cavity decreases with $h_{\text{eff}}$.   

Next we focus on the regime of high packing fraction. By increasing $\eta$ from the E state the particles show incipient positional order, forming smectic layers without changing their director field (not shown). Then, at higher packing fractions there is a complete structural change to the bridge smectic state (see Fig. \ref{fig12}c). In the Br$_{\text{S}}$ state the particles that were oriented along one diagonal in the E state rotate by $45^\circ$ generating three domains where the director is almost uniform. The domains are separated by domain walls where the director rotates by $90^\circ$ (the uniaxial order parameter vanishes at the domain walls, see Fig. \ref{fig12}c). The domain walls connect two corners and divide the cavity in three regions with uniform director. The size of the domains fluctuates but the central domain is always bigger than the others. The domain walls become more rigid as the density is increased. The same state has been predicted recently using density functional theory in a system of rectangles with restricted orientations (Zwanzig approximation) confined in the same geometry \cite{Miguel}. The authors of \cite{Miguel} classify the Br$_{\text{S}}$ state according to the number of smectic layers in the central domain. Such a criterion is not applicable in our case due to the large fluctuations of the domain walls, but obviously the number of smectic layers in the cavity varies with the side length. 

In order to estimate the packing fraction of the E-Br$_{\text{S}}$ transition we have made a histogram of the global tilt angle $\psi_g$ inside the cavity, i.e. the tilt angle resulting of a diagonalization of the tensorial order parameter formed by all the particles. In the isotropic state $\psi_g$ fluctuates between $0$ and $\pi$. As soon as the the elastic state arises, $\psi_g$ fluctuates between the values for both diagonal directions; the histogram shows two peaks at $\psi_g\approx\pi/4$ and $3\pi/4$. At high densities, but still in the E state, the system stops fluctuating between the diagonals (during the available simulation time) and the histogram shows only one peak either at $\psi_g\approx\pi/4$ or $3\pi/4$. Finally, in the Br$_{\text{S}}$ state there is a single peak centered at $\psi_g\approx0$ or $\pi/2$ governed by the particles of the main domain. Again at these packing fractions the particles cannot fluctuate between both equivalent states with $\psi_g\approx0$ or $\pi/2$ during the available simulation time. The behaviour of $\psi_g$ allows us to estimate the E-Br$_{\text{S}}$ transition as the packing fraction at which $\psi_g$ changes from $\psi_g\approx\pi/4$ or $3\pi/4$ to $0$ or $\pi/2$. The result is plotted in the state diagram, Fig. \ref{fig11}. The bigger the cavity is the higher the packing fraction at the E-Br$_{\text{S}}$ transition is. We can rationalize the transition as follows. Let $\Delta F$ be the excess in free energy of the confined system over a bulk undistorted state. In the Br$_{\text{S}}$ state $\Delta F_{\text{B}}=F_{\text{w}}+F_{\text{d}}$ with $F_{\text{w}}$ being the anchoring free energy due to the interaction with the walls and $F_{\text{d}}$ the contribution due to the domain walls. In the E state $\Delta F_{\text{E}}=F_{\text{w}}+F_{\text{e}}+F_{\text{c}}$, with $F_{\text{e}}$ accounting for the elastic deformations of the director field, and $F_{\text{c}}$ for the disclination cores. $F_{\text{w}}$ is similar in both cases because the anchoring is satisfied in both states. $F_{\text{d}}$ is proportional to the length of the domain walls and hence to $h_{\text{eff}}$. $F_{\text{c}}$ does not depend on the size of the cavity and, finally, the elastic energy is \cite{DF9582500019}  
\begin{equation}
F_{\text{e}}=\int_{\text{cavity}}d\vec r \left[k_1(\nabla\cdot\vec n)^2+k_3(\vec n\times(\nabla\times\vec n))^2\right],
\end{equation} 
where $\vec n(\vec r)$ is the director field and $k_1$ and $k_3$ are the splay and bend elastic constants, respectively. For rods confined in a circular cavity, the elastic energy grows logarithmically with the radius of the cavity~\cite{PhysRevE.79.061703}. Here, we have computed numerically the divergence and the rotational of the director in the E state and we have found that the dependence of the elastic energy with the cavity size is also weak, increasing slower than linear in $h_{\text{eff}}$. On the other hand, in the Br$_{\text{S}}$ state, the size of the domain walls is proportional to the size of the cavity, and hence $F_{\text{d}}\propto h_{\text{eff}}$. Therefore, for a fixed $\eta$ we expect the bridge state to be replaced by the elastic state at sufficiently big cavity sizes due to different dependence of $\Delta F$ with $h_{\text{eff}}$ in both states. The increase of the packing fraction at the  E-Br$_{\text{S}}$ can be understood given the behaviour of the elastic constants with the packing fraction; both $k_1$ and $k_3$ monotonically increase with $\eta$.  

In a very recent study~\cite{C4SM02087A} Garlea et al. have simulated a quasi-monolayer of hard spherocylinders confined in a square prism as well as the two-dimensional limit of discorectangles in a square cavity. The authors observe a state very similar to the elastic state in which the inner $-1/2$ defect and its adjacent corner defect form a kind of line defect. Actually, in our case, for the smaller cavities it is difficult to say whether those defects are actually two independent defects or whether they form a single structure. Garlea et al. have also found smectic ordering in their simulations, but in contrast to our findings they did not observe domain walls at high packing fractions, although they state in ~\cite{C4SM02087A} "... we do sometimes observe particles trapped perpendicularly to the smectic layers, invariably next to the wall". The differences at high packing fractions between both systems are probably due to the slightly different geometry of the particles (spherocylinders vs rectangles). In contrast to hard spherocylinders (or discorectangles in two dimensions), hard rectangles posses degenerate close packing states and have a higher tendency to cluster~\cite{raton064903,raton014501}. This may explain the presence of domain walls in a system of hard rectangles and its absence in a system of hard spherocylinders. It is unlikely that the dimensionality plays a dominant role because Garlea et al. have studied both the quasi two dimensional system and the strict two-dimensional limit, and found no differences between them.

\subsection{Square cavity with homeotropic walls}\label{sh}
\begin{figure}
\includegraphics[width=0.95\columnwidth,angle=0]{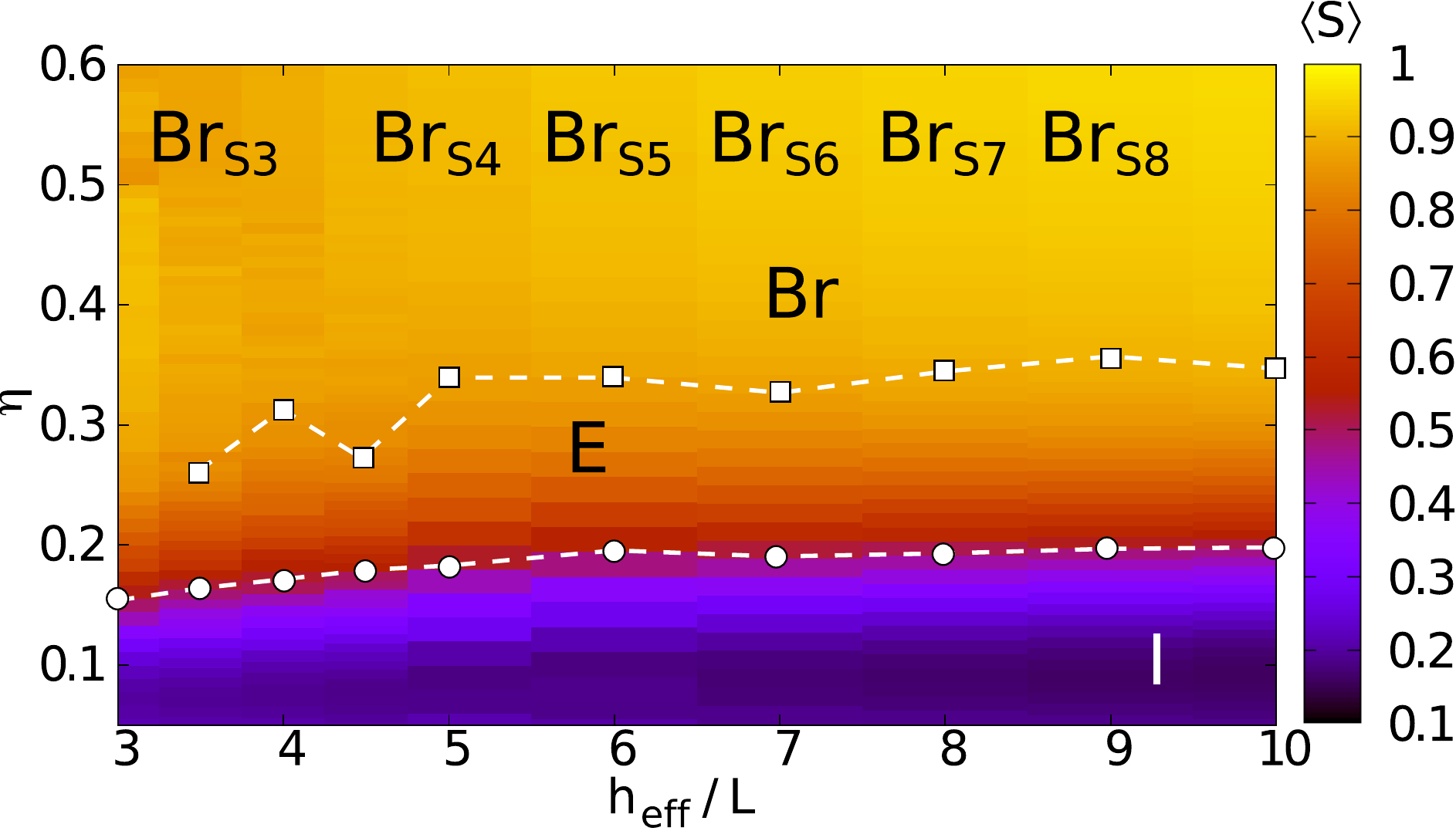}
\caption{Phase diagram of a fluid of hard rectangles ($L/D=20$ confined in a square cavity with homeotropic anchoring: packing fraction-side length plane. The color map indicates the average of the uniaxial order parameter $\langle S \rangle$. Empty circles mark the packing fraction at which $\langle S \rangle=0.5$. Empty squares roughly indicate the elastic-bridge nematic transition. Lines are guides to the eye.}
\label{fig13}
\end{figure}
\begin{figure*}
\includegraphics[width=1.0\textwidth,angle=0]{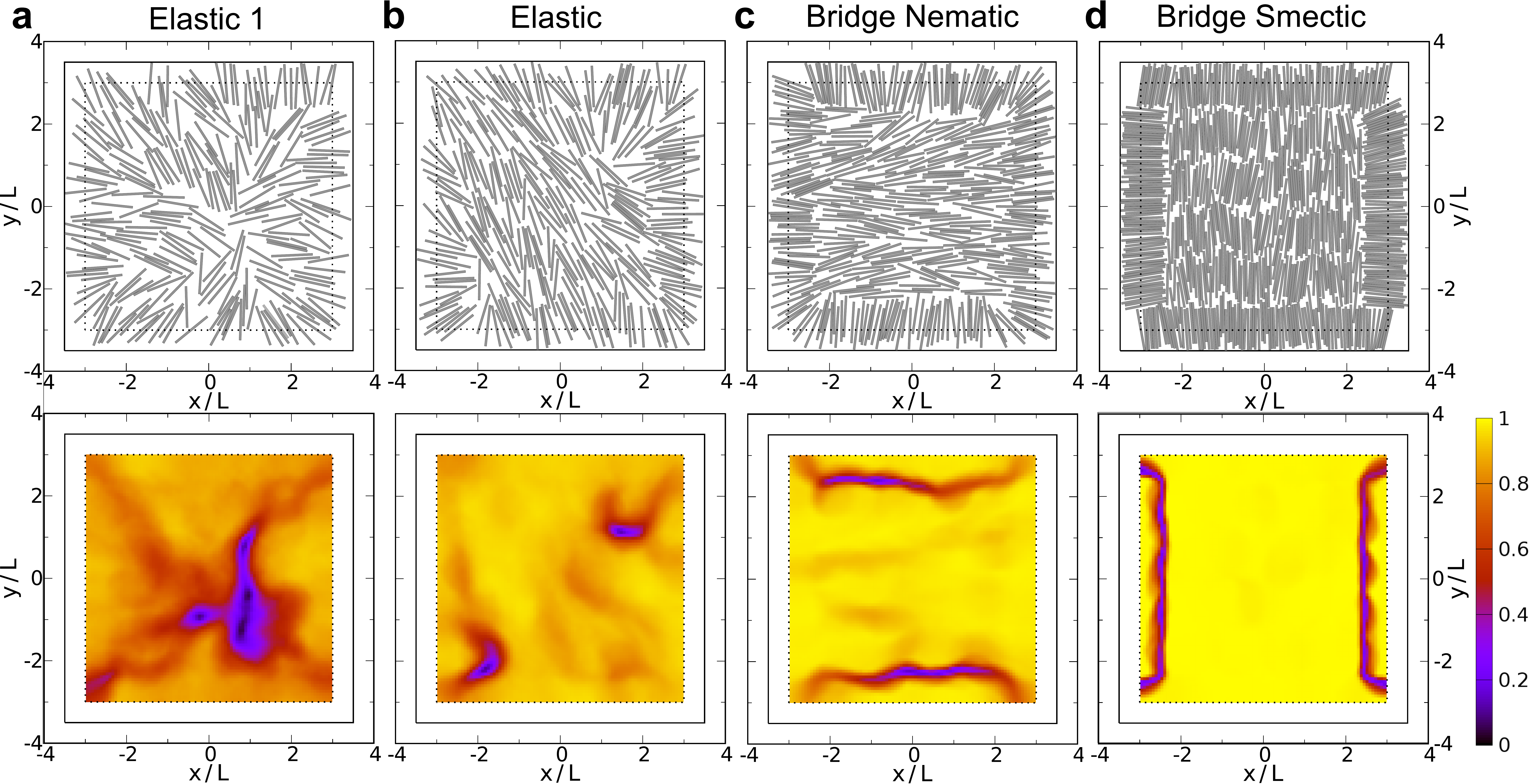}
\caption{Representative states in a square with homeotropic anchoring and side length $h_{\text{eff}}=7L$. The solid (dotted) square indicate the effective (actual) walls. Upper row: snapshots of the particles. Bottom row: local uniaxial order parameter. (a) single defect elastic state with $\eta=0.25$. (b) E state with $\eta\approx0.30$. (c) Br state with and $\eta\approx0.40$. (d) Br$_{\text{S}}$ state with and $\eta\approx0.60$.}
\label{fig14}
\end{figure*}

Finally we investigate the confinement of rods in a square cavity that promotes homeotropic anchoring (a schematic of the geometry is shown in Fig. \ref{fig1}d). The state diagram and representative states for $h_{\text{eff}}=7L$ are shown in Figs. \ref{fig13} and \ref{fig14}, respectively. Here, as in the case of the planar cavity, the elastic state consists of particles aligned along one diagonal (see Fig. \ref{fig14}b). However, in contrast to the planar cell, the alignment of the particles leads to the formation of only two disclinations with topological charge $+1/2$ (see the drop of the uniaxial order parameter in panel b of Fig. \ref{fig14}). The fluctuations in the position of the disclinations is high, much higher than in the planar cavity. This is most likely related to the dominant elastic deformations of the director involved in each disclination: splay-like deformations in the case of $+1/2$ disclinations and bend-like in $-1/2$ disclinations. As $k_3\ge k_1$ (see e.g., \cite{herasliqcrys}) we expect more fluctuations in the positions of $+1/2$ disclinations than in $-1/2$ disclinations. As an example of the high fluctuations of the $+1/2$ disclination cores, we show in Fig. \ref{fig14}a) a state where both disclinations have merged forming a single $+1$ disclination. This state is a variation of the elastic state that we observe sometimes, especially at low densities. This configuration is metastable because it involves higher elastic deformations and the energy of one $+1$ disclination core is higher than that of two $+1/2$ disclination cores (the energy of a disclination core increases with the square of its topological charge). 

By further increasing the density, we find a gradual transition from the elastic to the bridge nematic state (Fig. \ref{fig14}c). In the bridge nematic state there are three domains of particles with uniform director. In contrast to the planar cavity, here the bridge state is not accompanied by positional order because it appears at lower density (compare the position of the elastic-bridge transition in the states diagrams of Figs. \ref{fig11} and \ref{fig13}) and the particles remain in a nematic state. Another difference involves the domain walls that stay always at a distance of about one molecular length from the (effective) walls. The position of the domain walls fluctuates less than in the case of a planar cavity. At higher packing fractions the rods in the main domain form well-defined layers. We call this the bridge smectic state, B$_{\text{S}i}$, where $i$ indicates the number of layers of the main domain. The number of smectic layers is well defined due to the stable positions of domain walls in the cell. The number of layers is the result of commensuration between the side length of the cavity and the smectic layer spacing. The approximate regions of the distinct B$_{\text{S}i}$ states are indicated in the state diagram of Fig. \ref{fig12}. 

\section{Conclusions}
\label{conclusions}
In summary, we have performed a systematic analysis of the behaviour of two-dimensional hard rods confined in slit pores and in square cavities. In the case of slit pores we have shown that our simple hard core model contains much of the phenomenology observed in corresponding confined three dimensional systems. Examples are the capillary nematization and smectization in homeotropic pores, and the formation of linear and step states that occurs in hybrid planar cells. In addition we have found new states that have not been experimentally observed or theoretically predicted. An example is the smectic C and the brush state that we have observed in homeotropic cells. Both states break the anchoring imposed by the surfaces. The asymmetric brush state breaks also the symmetry of the cell. In all cases we have rationalized the stability by comparing the excess in free energy to the corresponding undistorted bulk phase. 

In recent experiments on vertically vibrated monolayers of rods confined in a circular cavity \cite{Galanis2006,Galanis2010a} the same textures were found as MC studies predict for equilibrium rods \cite{domainwalls}. Actually, the elastic state we have found in the square planar cavity has been observed in vibrated granular rods \cite{Galanis2006}. Granular materials flow and diffuse anomalously \cite{Yadav,Aranson1}. Although being non-thermal fluids, under certain circumstances such systems form steady states with the textures of thermal fluids. A comparison between MC simulation of confined rods (thermal fluid) and vibrated granular rods (non-thermal fluid) would help to find the analogies between both systems. The elastic state of the planar square cell has been also observed experimentally in confined actin filaments \cite{Mulder}, colloidal particles \cite{C4SM01123F}, and predicted using Onsager-like density functional theory \cite{Chen}. The authors of \cite{C4SM01123F} found, using experiments on confined colloids and Oseen--Frank elastic theory, that the elastic state is metastable with respect to another state that contains two corner disclinations and is free of bulk disclinations (diagonal state). In the diagonal state the total deformation of the director is higher than in the elastic state, but on the other hand, the diagonal state has no bulk disclinations. The total elastic energy decreases with the size of the cavity, and the energetic cost associated to a disclination is independent of cavity size. Hence, we expect the diagonal state to replace the elastic state in our system for cavities much bigger than the ones studied here. The rate between the splay and bend elastic constant also plays a role determining the relative stability of the confined states. In \cite{C4SM01123F} the case of equal elastic constants is analysed, whereas in our system we expect the bend elastic constant to be much higher than the splay one. 

Although we have analysed a two-dimensional model, our results may be of relevance to gaining a better understanding of three-dimensional systems where similar phenomenology has been already found. For example, capillary nematization \cite{heras:4949,0953-8984-19-32-326103} and smectization \cite{PhysRevLett.94.017801,PhysRevE.74.011709,0953-8984-22-17-175002} have been studied in confined rods and platelets between two parallel walls. The hybrid cell has been also analysed in three dimensions \cite{ponce}, and phases with the same symmetry as those found here appear. Our results indicate that other states, not observed yet in three-dimensional systems, can arise under extreme confinement. For example, states that break the anchoring, like the smectic C or the brush states found here, or states with symmetry breaking (i.e., asymmetric states in confined symmetric pores) such as the asymmetric brush state. Those states might be difficult to find in e.g. density functional studies where one typically assumes that the symmetry of the order parameter profiles is the same as the one imposed by the surfaces.  

It is interesting to compare the confinement of rods in square cavities, Secs. \ref{sp} and \ref{sh}, with the recent study of confined rods in circular cavities \cite{domainwalls}. In both cases at high densities the system form domain walls in an attempt to reduce the elastic distortions of the director field. Although the domain walls will probably disappear in larger cavities, they might be a general mechanism to reduce elastic stresses under extreme confinement. 

Some of the states found in the slit-pore geometry show lateral ordering, such as for example the brush smectic states. For selected pore sizes, we have performed simulations varying the lateral size of the cell, $h_y$ from $10$ to $20L$ and no differences have been found. We are, therefore, confident that the lateral ordering is not induced by the applied boundary conditions. Nevertheless, monte carlo simulations in the isothermal--isobaric ensemble (NPT) might help to elucidate the role that the lateral size of the pore plays in the stability of such states.

We have restricted the analysis to hard rectangles with length-to-width ratio of $20$. We expect a similar phenomenology for particles with aspect ratio higher than $\sim7$ because the bulk behaviour is qualitatively the same. However, for particles with shorter aspect ratios completely new phenomenology will presumably appear because states with tetratic correlations are stable in bulk \cite{VelascoSPT,Raton2006} and might modify the phase behaviour presented here.

\section{Acknowledgments and author contributions}
We thank E. Velasco for useful discussions.\\

T. Geigenfeind and S. Rosenzweig contributed equally to this work.

\end{document}